\documentclass[manuscript]{aastex}
\usepackage{natbib}
\usepackage{graphicx}
\usepackage{dcolumn}
\usepackage{bm}
\usepackage{multirow}
\usepackage{amsmath}
\usepackage{amsfonts}

\shorttitle{N~V, O~III] and He~II in TTSs magnetospheres}
\shortauthors{A.I. G\'omez de Castro}

\begin{document}

\title{Evidence of hot high velocity photoionized plasma falling  on
actively accreting T-Tauri Stars.}
\author{Ana Ines G\'omez de Castro}
\affil{Grupo de Investigaci\'on Complutense AEGORA and S.D. Astronom\'ia y Geodesia, 
Fac. de CC Matem\'{a}ticas, Universidad Complutense, 28040 Madrid, Spain}
\begin{abstract}

The HeII(1640\AA ) line and the resonance doublet of NV(UV1)  provide a good diagnostic tool to constrain the excitation 
mechanism of hot ($T_e >40,000$K) atmospheric/magnetospheric plasmas in T Tauri Stars. Making use of the data available 
in the  Hubble Space Telescope Archive, this work shows that there are, at least, two distinct physical components 
contributing to the radiation in these tracers: the accretion flow sliding on the magnetosphere and the atmosphere.

The NV profiles are symmetric and at rest with respect to the star, in most sources.  The velocity dispersion of the 
profile increases from non-accreting ($\sigma = 40$~km~s$^{-1}$) to accreting 
($\sigma = 120$~km~s$^{-1}$) TTSs suggesting that the macroturbulence field in the line formation region decreases 
as the stars approach the main sequence. 
Evidence of the NV line being formed in a hot solar-like wind has been found in RW Aur, HN Tau and AA Tau.

The HeII profile has a strong narrow component that dominates the line flux; the dispersion of this component ranges 
from 20 to 60 km~s$^{-1}$ . The source of this radiation is uncertain. Current data suggest that both accretion shocks 
and atmospheric emission might contribute to the line flux.   In some sources, the He II line shows a broad and redwards 
shifted emission component often accompanied by semiforbidden O~III] emission
that has a critical electron density of  $\sim 3.4 \times 10^{10}$~cm$^{−3}$.

In spite of the different origins (inferred from the kinematics of the line formation region), NV and HeII fluxes 
are strongly correlated, with the only possible exception of some of the heaviest accretors.

\end{abstract}

\section{Introduction}

Solar-like pre-main sequence (PMS) stars and, in general, the low mass PMS stars (M$_* < 2 M_{\odot}$), or T~Tauri stars (TTSs), are complex dynamical systems made of two basic components, star and accretion disk, as well as a dynamical interface,
the stellar magnetosphere (see G\'omez de Castro 2013, for a recent review).  Magnetic fields of few kG have been detected in the surface of the TTSs (Guenther et al 1999; Johns-Krull et al 1999; Johns-Krull 2007). The surface field is not bipolar but it has a rather complex structure as the Sun's field has (Johns-Krull et al 2004). Hence, though higher order, multi-polar components fall off more rapidly with radius than the dipolar field,  it is   difficult to track the final path followed by matter from the inner disk border to the stellar surface. In addition, the magnetosphere has its own dynamics and forcing due to the interaction with the disk. 

Unfortunately, the characteristics of the TTSs extended atmosphere and magnetospheres are still escaping the diagnosis (see Hartmann 2009). Little is known about them, apart from having a density of $\sim10^9-10^{11}$~cm$^{-3}$ and an electron temperature between some few thousand Kelvin and 100,000~K (G\'omez de Castro \& Verdugo 2007, hereafter GdCV2007).  Recent attempts to derive the magnetospheric properties from optical lines have shown that, for instance, the H$\alpha$ profile is strongly dependent on densities and temperatures assumed inside the magnetosphere and in the disk wind reagion (Lima et al. 2010).  The line widths of typical atmospheric/magnetospheric tracers are about 200-300 km/s that exceed by far what expected from thermal or rotational broadening even if the lines are assumed to be formed in a magnetosphere that extends to some 4-5 stellar radii and corotates with the star.
As today, it is still unclear whether the broadening is produced by unresolved macroscopic flows or by magnetic waves propagating on the magnetospheric field (Hartmann et al. 1982) .  In general, the broadening of magnetospheric tracers does not vary significantly in time, pointing out that the average motions are rather stable.
The combined effect of funnel flows and inclined magnetic rotators as simulated by Romanova et al (2004,2012) is the current baseline for the numerical simulation of TTSs magnetospheres. The stellar field is assumed to be anchored in the inner part of the disk creating a sheared layer between the rigid body rotation of the star and the Keplerian rotation of the disk. This interaction has a profound influence on the star and the accretion flow but also acts as a dynamo that transform part of the angular momentum excess in the inner disk into magnetic field amplification that self-regulates through quiescent periods of field building-up and eruptions when the energy excess is released (see G\'omez de Castro \& von Rekowsky 2011, for an evaluation of the UV output from such interface). However, magnetospheric heating processes are poorly known and accretion shock models are unable to predict the observed line fluxes/broadening (Johns-Krull 2009).

The atmospheric and magnetospheric energy output is released mainly in the ultraviolet. The richness of spectral tracers for a broad range of magnetospheric temperatures and densities is unmatched by any other range. There are several studies of the atmospheric/magnetospheric properties of the TTSs based on low dispersion UV data (Lemmens et al. 1992, Hu\'elamo et al. 1998, Johns-Krull et al. 2000, Yang et al. 2012, G\'omez de Castro \& Marcos-Arenal, 2012, hereafter GdCMA2012). However, only the Cosmic Origins Spectrograph (COS) (see Green et al 2012 for a description of the instrument), on board the Hubble Space Telescope (HST), has been sensitive enough to gather high signal-to-noise ratio (SNR) profiles of hot plasma tracers such as the N~V [UV1] resonance multiplet or the He~II H$\alpha$ transition with a resolution above 10,000 in late K and M-type TTSs (see Penston \& Lago 1984 for C~IV profiles obtained with the IUE, Ardila et al 2002, for HST profiles obtained with the Goddard High Resolution Spectrograph and Ayres 2005, for the CoolCat set based on HST data obtained with the Space Telescope Imaging Spectrograph and compare them with the presented in Sect.~2). 

Following a previous work where the magnetospheric properties of TTSs were examined based on low dispersion HST data (GdCMA2012), the high resolution profiles of the N~V [UV1], O~III] and He~II(1640 \AA) transitions are analysed in this work with two main objectives. Firstly, it is intended to determine whether it is feasible to discriminate the contribution to the high energy radiation flux from the high density atmospheric plasma and from the accretion flow. 
Moreover, the unexpected correlation between X-ray flux and the high energy UV tracers found by GdCMA2012 is re-examined on the light of the kinematic information contained in the high resolution HST/COS profiles. 
In Sect.~2 the observations are described. The results are presented in Sect.~3. Two components are found to contribute to the He~II flux a Low Density Component (LDC) associated with the accretion flow and a High Density Component (HDC) of more uncertain origin. The discussion on the possible source of the HDC, its association with accretion shocks and the properties of the LDC are addressed in Sect.~4. The article concludes with a brief summary on the relevance of these results.

\section{Hubble Space Telescope (HST) observations}

The He~II profiles of TTSs in the Taurus-Aurigae star forming region have been retrieved from the HST archive. Only high resolution observations have been considered. Most of them have been obtained with the Cosmic Origins Spectrograph (COS) and the gratings G130M and G160M. The resolution is $\sim$24,000 and  each target has been observed three times with slight offsets in the wavelength range to guarantee that the 18.1\AA\ gap between the two segments in the FUV detector is covered (see COS Handbook). 
Moreover, some few TTSs have been observed with the Space Telescope Imaging Spectrograph (STIS), namely, T~Tau, DR~Tau, and DF~Tau. Only the T~Tau profile had a high enough signal-to-noise ratio (SNR) in the He~II line to be considered for this work. The log of the observations is provided in Table 1; additional information on the HST programs that obtained these observations (programs IDs. 11533, 11616 and 8627) can also be found in the Table. 

The observational strategy allowed to search for variations in the profiles in time scales of $\sim 20$ minutes but neither flux, nor morphology were found to display significant changes. The one-dimensional spectra produced by the COS calibration pipeline (CALCOS v2.17.3) were aligned and co-added\footnote{Note that there are small wavelength windows in the spectra without flux measurements. This effect has been taken into account in the calculation of the average spectra  from the, typically, three observations obtained per star.}. COS targets must be centred to within $0.1-0.2$ arcsec to achieve the nominal wavelength accuracy of $\pm 15$~km~s$^{-1}$. The  (R(3)~1-7) 1489.636~\AA\ and (P(5)~1-7) 1504.845~\AA\  H$_2$ lines have been used to set the zero of the wavelength scale for the targets in Table~1. H$_2$ emission is dominated by the molecular disk around the stars in most sources (France et al. 2012). The lines have been selected to be strong (from Herczeg et al. 2002) and detectable in most of the sources. The profiles of the 1489.636~\AA\ line are plotted in Figure~1 (see also Figure~3 in France et al.~2012) and the He~II profiles are plotted in Figure~2. No shifts have been applied to DN~Tau and IP~Tau observations because the H$_2$ emission was too weak to be used for this purpose. Neither shifts have been applied to HBC~427, LkCa~19 and LkCa~4 because H$_2$ emission was not detected. The H$_2$ profiles are sometimes asymmetric with respect to the rest wavelength, very especially in RW~Aur. In this case, the original zero from the CALCOS pipeline has been left. Only a subset of AA~Tau observations has been used since guide star acquisition failed (see  Fig.~3). As a result, the first two exposures produced similar profiles while the last one produces a slightly broader and more red-shifted (by $\sim 0.1$\AA\ or 18~km~s$^{-1}$ at 1640 \AA ) profile. For this star, the last observation was rejected and only the two first observations were averaged to produce the profiles in Figs.~1 and 2. 

The He~II profiles can be generically described as composed of a bright and narrow emission feature and a broad, weaker component that differs from one star to another. Notice that the He~II lines are very strong; this fact, together with the strong H$_2$ emission, contributes to the continuum jump in the low resolution Advanced Camera System on HST reported by GdCMA2012. 

Close to the He~II line, there are the O~III]$_{1665}$ intercombination lines; a doublet with components at $\lambda \lambda 1660.802$ and 1666.156, which originate under transitions from the level 2s2p$^3$ $^5$S$_2$ to the term 2s$^2$2p$^2$ $^3$P with J=1 and J=2, respectively. The components should have an intensity ratio $\simeq$1:3; equal to the ratio of their transition probabilities (145~s$^{-1}$ and 426~s$^{-1}$). The transition is optically thin to a critical density of  $3.4 \times 10^{10}$~cm$^{-3}$.  The O~III] profiles are represented in Fig.~4. The two lines of the multiplet are observed in GM~Aur, DF~Tau, HN~Tau, DR~Tau, SU~Aur, RW~Aur and DE~Tau; however the flux of the weakest, 1660.802~\AA\ line, has only been measured for strong sources. In all cases, the flux ratio between the two lines of the multiplet is 1:3 (within the error bars).

To complete the view on the distribution of hot plasma in the TTSs environment, also the profiles of the resonance UV~1 multiplet of the N~V have been retrieved from the HST archive (see Table~2). In the blue edge of the 1238.8~\AA\ line, the strongest in the doublet, there are some  narrow emission lines produced by molecular hydrogen (lines: $\lambda 1237.918$\AA\ 1-2 P(8), $\lambda 1237.589$\AA\ 2-2 R(11)) that somewhat blur the profile.   The zero of the wavelength scale has been set again, resourcing to H$_2$ emission lines. The (P(2)~0-4) 1338.63~\AA\ line has been used for this purpose, since it is strong in most of the stars and it is not blended with other features (see Figure~5 with the H$_2$ profiles).  As mentioned above, the original zero of the wavelength scale has not been shifted for DN~Tau,  IP~Tau, HBC~427, LkCa~19 and LkCa~4  because the H$_2$ lines were either very weak or absent. The N~V profiles can be most generally described by a single component that ranges from being narrow in stars like HBC~427, LkCa~19 and LkCa~4 to be broad and asymmetric  in AA~Tau or GM~Aur (see Fig.~6). 

Some relevant properties of the TTSs to be used in Sect.~3 are gathered in Table~3. Notice that there are wide variations in published values of important parameters such as the stellar luminosity or the extinction  (see also comments in GdCMA2012); data in Table~3 are gathered for reference for other researchers.  The X-ray fluxes have been retrieved from the XMM-Newton extended survey of the Taurus molecular cloud (XEST) (Guedel et al. 2007). The He~II, the 1238.82~\AA\ N~V and the  O~III] lines fluxes have been measured after subtracting the local continuum. Also the fluxes for the  1489.636 and 1338.63 H$_2$ lines have measured (see Sect.~5).The lines fluxes are provided in Table~4; they are not extinction corrected.  For some sources, the measurement of the 1238.82~\AA\ N~V flux has required subtracting the nearby H$_2$ features. In such cases, the H$_2$ flux has been subtracted by linear interpolation in the  N~V profile. However, there are some few profiles where this linear interpolation was uncertain  (see quality flags in Table 4).

\section{Results}

From figures 1 to 6 a generic trend can be inferred:

\begin{itemize}

\item The {\it weak-line TTSs (WTTSs)} in the sample, {\it i.e.} evolved TTSs with no evidence of mass infall (LkCa~19, 
LkCa~4, HBC~427), do not  display H$_2$ emission,  neither nebular O~III] emission. They have only rather narrow He~II 
and N~V lines.

\item {\it Intermediate objects (TTSs)} like IP~Tau, DN~Tau and V836~Tau have weak H$_2$ and no nebular O~III] 
emission. Both He~II and N~V lines have narrow emission profiles.

\item {\it The Classical TTSs (CTTSs)}   emit in all these tracers (see below).

\end{itemize}

CTTSs cover a broad range of profiles morphologies. Strong H$_2$ emission is detected in all of them and the lines are 
narrow unless in RW~Aur~A (see France et al 2012). Nebular O~III] emission is detected in all of them unless in DM~Tau, 
UX~Tau~A and AA~Tau. O~III] is especially strong in T~Tau, DF~Tau, HN~Tau, DR~Tau and RW~Aur.  SU~Aur, and  DE~Tau 
seem to be intermediate objects. Hints of possible O~III] emission at 1666~\AA\ are seen in GM~Aur, unfortunately 
the S/N is low and the weakest component of the multiplet is not detected. A low S/N feature is  detected at 1666~AA 
in DN~Tau, IP~Tau, AA~Tau and  UX~Tau spectra. Observations of the C~III] intercombination transition line are only 
available for three of the stars in the sample, namely, DE~Tau, T~Tau and RW~Aur and no significant differences have 
been found between the O~III] and the C~III] profiles obtained either with the Goddard High Resolution Spectrograph 
(GHRS) (compare  Fig.~4 with Fig.1 in G\'omez de Castro et al 2001) and STIS (compare  Fig.~4 with Fig.1 in 
G\'omez de Castro et al 2003).  

In general, the He~II profiles can be described by a narrow emission component superimposed on a broader contribution 
that mimics the observed in the O~III] nebular lines, when high enough S/N data profiles are observed. 
This effect is clearly apparent in  HN~Tau and RW~Aur. Note that in DR~Tau and DF~Tau, the overlap is lost at 
bluewards shifted velocities; there is bluewards shifted O~III] emission with no He~II counterpart.  This bluewards
shifted excess could be caused by the contribution of an unresolved jet to the line emission. As shown for RY~Tau by
GdCV2003, the semiforbidden emission has two components: one associated with the jet and another
with the accretion flow that were disentangled for this source due to their variability. The N~V profiles however, do not follow the same morphological trend. 
A rather narrow symmetric profile is observed in DE~Tau that becomes wider in DM~Tau, DF~Tau, UX~Tau~A, GM~Aur 
and SU~Aur, all of them sources with absent or weak O~III] nebular emission. DR~Tau, HN~Tau, RW~Aur and AA~Tau 
display very peculiar profiles that clearly indicate that N~V emission is not produced in the stellar atmosphere 
but in another dynamical component. In particular, the N~V profiles of RW~Aur, HN~Tau and AA~Tau are asymmetric 
extending from red-wards shifted velocities to peak at blue-wards shifted velocities; moreover, the RW~Aur profile 
peaks at the velocity of the optical jet (Hirth et al 1997) that was also detected in the C~III] intercombination 
line by  G\'omez de Castro \& Verdugo, 2003. This  type of profile asymmetry has also been detected in the C~III] 
profile of RY~Tau (GdCV2007), who pointed out that the line could be formed in a pre-main sequence analogue of the 
Solar wind. In this context, it is worth remarking that the H$_2$ profile of RW~Aur has a completely different 
asymmetry, the flux peaks to the red of the line, suggesting the presence of infalling cold molecular gas similar 
to that observed in the pre-main sequence close binary AK~Sco (see also G\'omez de Castro et al 2013). 

These groups can be cleanly recognised in velocity dispersion diagrams. The characterisation of the underlying velocity field and thermal properties of the line emission region is complex in the TTSs environment. Instead of using the standard fitting to Gaussian or Voight profiles, it is preferable to characterise the profile in terms of the standard Pearson statistics moments  and measures, i.e., mean or centroid, dispersion, kurtosis and skewness. They provide a quantitative measurement of the deviation of the profile from the expected for a thermal plasmas; a normal distribution convolved with the Line Spread Function (LSF) of COS (Kriss 2011). 
Note that in this approach, the profile is assumed to be formed by the contribution to the line flux of independent gas parcels, {\it i.e.} the (background subtracted) profile is treated as a histogram of the flux emitted per parcel in the wavelength (velocity) space; a similar approach was followed by GdCV2007 to compare the observed C~III] and Si~III] profiles of RY~Tau with the theoretical predictions. This treatment permits to characterise the profiles that are formed in complex velocity fields. In Table~5, the dispersions of the He~II and N~V lines are provided together with that obtained for two control lines, the H$_2$ transitions at 1339\AA\ and 1489\AA ; RW~Aur, HN~Tau and AA~Tau are not included because their N~V profiles are peculiar. As shown in the bottom panel of Fig.~8, the average dispersion of the H$_2$ profiles is $31\pm 5$~km~s$^{-1}$ and $43\pm 13$~km~s$^{-1}$ for the 1339\AA\ and 1489\AA\ lines, respectively.
There is not any correlation between the broadening of these two H$_2$ lines, pointing out that the scattering of the dispersions in the diagram  is related with random effects associated with the measurement process. A different trend is drawn from the N~V and He~II dispersions. The dispersions of the WTTSs and DN~Tau are comparable to those measured in the H$_2$ line. In intermediate objects, like V836Tau and IP~Tau, the dispersions in the He~II line are comparable to those measured in the H$_2$ lines but the N~V lines are significantly broader.  Finally in the CTTSs both $\sigma (N~V)$ and $\sigma (He~II)$ are larger than $\sigma (H_2)$.  Note that stellar rotation may contribute to the line dispersion; SU~Aur, the fastest rotator in the sample has also the largest dispersion. However, a large dispersion can also be produced by the profile asymmetry. For instance, DM~Tau has dispersions comparable to those measured in SU~Aur and it is one of the slowest rotators in the sample (see Table~3).  Two objects do not follow this trend: DR~Tau and DE~Tau, both have intermediate dispersions in the N~V lines and large dispersions in the He~II lines caused by the broad emission component. A quick inspection in the summary of the TTSs properties (see Table~3) indicates that the only possible cause of this discrepancy is the high accretion rate, as otherwise, expected.

\subsection{Two hot plasma components in the TTSs}

From figures 2, 4 and 6, it is clearly inferred that there are, at least, two different plasmas contributing to the spectral lines under study:

\begin{itemize}

\item {\it A low density component (LDC)} that it is most conspicuously traced by the O~III] line. The LDC also produces the He~II broad emission component observed in RW~Aur, HN~Tau, DR~Tau and DF~Tau. The critical density of the O~III] sets an upper limit to the electron density\footnote{ Note that for electron densities above the critical density, there may still be line emission though it damps rapidly.}  of the plasma in the LDC of $\simeq 3.4 \times 10^{10}$cm$^{-3}$,
(see also Sect.~4.2). The LDC profiles display a non-thermal broadening  and draw a complex velocity field around the stars, i.e., it is not associated to a simple standing atmospheric structure whose kinematics is dominated by stellar rotation. In fact, the LDC profiles seem to rather trace  some kind of complex magnetospheric infalling pattern, high above the stellar surface. Also, in some cases, could be associated to unresolved wind structures (G\'omez de Castro \& Ferro-Font\'an 2005), as the reported for RY~Tau (GdCV2007). Notice that the co-existence of O~III] and He~II radiation from the same kinematic structure  would point out an unrealistically high electron temperatures for the line emission region ($\log T_e (K) \sim 5.4$), if collisional equilibrium at a single temperature is assumed  and electron densities below the O~III] critical density are considered\footnote{Calculations made using the Chianti data base: www.chiantidatabase.org.}.    In fact, its UV spectrum is reminiscent of that observed in photoionized nebulae (see also Sect.~4). To the current sensitivity, the contribution of this component to the N~V flux is negligible. 

\item {\it A high density/temperature component (HDC)} that dominates the N~V emission. O~III] profiles are very different from N~V profiles
suggesting that the density is of the N~V formation region is higher than the O~III] critical density.

\end{itemize}

Though the kinematics of the N~V emission region is clearly different from that traced by the LDC, the N~V flux is 
correlated  with the He~II flux as shown in Fig.~9;  the Spearman rank correlation coefficient is $r_s = 0.87$ (with significance level, $\alpha = 0.001$, see Sachs 1982 for details) and, \\
$$\log \left(F(HeII)\right)  = (0.8 \pm 0.1) \log \left( F(NV) \right) - (2.5 \pm 1.7)$$ 
\noindent 
with RMS = 0.29 (see bottom panel in Fig.~9). Also the fluxes normalised to the stellar surface are correlated
with $r_s = 0.82$,  with $\alpha = 0.002$ (see top panel in Fig.~9) and,\\
$$\log \left( \frac{F(L(HeII)}{F_{\rm bol}} \right)  = (0.9 \pm 0.1) \log \left(  \frac{F(NV)}{F_{\rm bol}} \right) + (0.0 \pm 0.6)$$ 
\noindent 
with an RMS=0.31. The normalised flux  is defined as the rate $F_{He II}/F_{\rm bol,*}$ or $F_{N V}/F_{\rm bol,*}$
and was introduced by GdCMA2021 to provide a measure of the line emissivity weighted over an unknown thickness but corrected from stellar radii and surface temperature. In this manner, the  normalised fluxes compensate for scaling effects associated with the broad range of mass, luminosity and stellar radius covered by the TTSs. Stars whose He~II flux
has a significant contribution from the LDC are marked in the plot. Notice that they are evenly distributed
in the figure suggesting that He~II and N~V fluxes are correlated, independently of whether the He~II flux is dominated by the narrow
emission component.

\subsection{The connection between UV and the X-ray radiation from the TTSs}

Based on low resolution observations, GdCMA2012 pointed out that the normalised  He~II flux anti correlates with the strength of the X-ray flux as derived in the XEST survey (Guedel et al 2007) carried out with the XMM-Newton telescope in the 0.3-10.0~keV band.  In Fig.~10, the normalised He~II fluxes from the high resolution COS/HST observation  (see Table~3) are represented against the normalised X-ray luminosities as derived from the XEST survey. Also the Chandra/ACIS observations of LkCa~4, DE~Tau, GM~Aur from Yang et al. (2012) are used; they are integrated X-ray luminosities in the 0.3-10~keV range. The He~II flux has been extinction corrected according to Valencic et al (2004) assuming R=3.1 and extinctions in Table~3. The low dispersion, GdCMA2012 
data, have been also plotted for those sources with no available high dispersion data. There is significant contribution to the He~II flux from the LDC in some few sources: SU~Aur, HN~Tau, DE~Tau and DM Tau, that are marked in the figure.
A first inspection of the Fig~10, shows three groups. The WTTSs (HBC~427, LkCa~4, HD283472), UX~Tau and the fast rotator
SU~Aur are very close to the main sequence stars regression line. The CTTSs follow the generic trend pointed out 
by GdCMA2012, the He~II flux increases as the X-ray flux decreases. Notice that the sources with strong LDCs are far from 
this trend, excluding the fast rotator SU~Aur. If  HN~Tau, DE~Tau and DM Tau are excluded, the Spearman rank correlation coefficient is $r_s=-0.591$,
with $\alpha = 0.033$,  and the least square fit is, 
$$\log \left( \frac{F(L_X)}{F_{\rm bol}} \right)  = (-0.20 \pm 0.05) \log \left(  \frac{F(HeII)}{F_{\rm bol}} \right) - (4.3 \pm 0.2)$$ 
\noindent 
with RMS = 0.15. Uncertainties in the A$_V$ values may produce a shift towards large  $f_{He II}$  in the plot reinforcing the trend; note that the regression line runs nearly parallel to the extinction direction. To examine its effect, the same diagram has been plotted in the bottom panel of Fig~10 using, stellar luminosities, X-ray luminosities and A$_V$ values from the recent compilation by Yang et al (2012). As shown, the three groups are no so cleanly separated and the 
regression line is more clear; only DE~Tau is far from the trend  and has been excluded from the calculation. 
The results are similar though the negative slope is softer: 
$$\log \left( \frac{F(L_X)}{F_{\rm bol}} \right)  = (-0.17 \pm 0.07) \log \left(  \frac{F(HeII)}{F_{\rm bol}} \right) - (4.1 \pm 0.1)$$
\noindent 
with RMS = 0.15 and   $r_s = -0.561$, with $\alpha = 0.036$.  

In summary, though the clean separation between WTTSs and accreting TTSs may be an extinction associated effect, the trend
to release preferentially the high energy excess in the UV rather than in the X-ray channel in accreting objects holds with the only possible exception of the sources with a strong contribution the He~II flux from the LDC. 

From the current data sets and observations, it cannot be ascertain whether there is a statistically meaningful deviation of the sources with strong LDCs from the main trend.
Unfortunately, there are not X-ray measurements of DR~Tau or DF~Tau and  RW~Aur was only detected to have a low soft X-ray flux with the EINSTEIN satellite (Damiani et al 1995). 
If present, such a trend could indicate that X-ray radiation is dominated by different components in sources with strong LDCs (strong nebular component) and in sources with
weak or absent LDCs. The X-rays energy distribution of the TTSs is often modelled by two components: a soft component at $T_s \simeq (2-5)~10^6$~K and a hard component at $T_h \simeq (1.5-3)~10^7$~K  (see i.e. Glassgold et al, 2000).  The hard X-ray component is thought to be associated with magnetic energy
release in the stellar coronae. The nature of the soft X-ray component is more uncertain and often, it has  been hypothesised that could be formed in accretion shocks (Lamzin 1998, Gullbring et al 1998). Unfortunately,  only three stars in our sample namely, T~Tau, SU~Aur and HBC~427 have a high enough count rate to allow a spectral fitting to two different optically thin plasmas and non-conclusive results could
be derived from the fits (see Table~6 with the two-components fit to the X-ray spectrum of these sources (from Table~6 in Guedel et al. (2007)). The  X-ray spectrum of SU~Aur, a CTTS, is dominated by the low temperature component with T=5.22~MK however, both soft and hard X-ray component, have similar emission measures in HBC~427,a non-accreting WTTSs. Moreover, the hard X-ray component dominates the X-ray spectrum of the CTTS, T~Tau.

\section{Discussion}

TTSs are complex objects; they are  convective PMS stars where a solar-like dynamo begins to set in, while still the fossil magnetic field is diffusing. TTSs are surrounded by an external dynamo that powers and makes rise the stellar magnetosphere to the inner border of the molecular disk (see Romanova et al 2012 for recent simulations). Matter from the disk, slides down onto the star along the magnetospheric field lines to end, free-falling onto the open holes of the magnetic configuration. In this environment, hot plasma radiating in the UV tracers studied in this work can be located in the magnetosphere, in the atmosphere, in the accretion shocks and also in the outflow (either solar-like or driven from the star-disk magnetic interface or the disk). Both, magnetosphere and outflow, have significantly lower densities than the stellar atmosphere or the accretion shocks; as a result, spectral lines radiation is dominated by radiative de-excitation processes and forbidden and semiforbidden transitions are strong from this plasma. In the dense atmosphere, collisional de-excitation is relevant and forbidden transitions are quenched. UV semiforbidden transitions cannot be observed from the accretion shock itself, because it is too hot and dense however, the soft X-ray radiation produced in the shock front photoionizes the preshock gas, which has a 
density similar to that of the stellar magnetosphere and may produce forbidden lines radiation (G\'omez de Castro \& Lamzin, 1999). Unfortunately, the high column density prevents the UV radiation from the photoionization cascade to escape easily from the accretion column. Also, the profiles of some tracers, from the infrared He I transition (Beristain et al. 2001, Fisher et al. 2008) to the UV lines do not agree with the predictions of accretion  shock models (Johns-Krull 2009). Within this context, the data presented in Sect.~3 provide some amazing results:

\begin{itemize}

\item  The He~II, O~III] and N~V fluxes do not depend on the spectral type. This confirms that {\it the line emission is not dominated by main sequence like atmospheric magnetic activity}, {\it i.e.} with the release of the magnetic energy produced by the stellar dynamo, since this one depends on the spectral type (see e.g. Ayres et al 1995 and GdCM2012). 

\item The He~II and N~V fluxes  correlate well independently of whether the line profiles are very different, {\it e.g.}, independently of whether the line emission is produced in the same physical structure. This confirms that all processes 
(accretion, atmospheric emission and outflow) are coupled, as otherwise expected  (see Gomez de Castro 2013 for a recent review). 
In turn, it makes difficult to get specific tracers of individual processes without the kinematical information, {\it i.e.}
without high resolution spectroscopy.

\item The high resolution profiles of the  N~V  line show a symmetric line broadening that increases from non accreting to accreting stars, being significantly suprathermal in these last sources (see Table~5).  The profile shape and the density of the line formation region suggests an atmospheric origin (with the exceptions already mentioned in Sect.~3).  The connection between line broadening and accretion suggests that
the density and extent of the high atmospheric layers depends on the accretion rate, {\it i.e.} on the evolutionary state, as otherwise predicted from the theoretical models (D'Antona \& Mazzitelli 1997, Siess et al 2000). Transport of magnetic energy from the stellar interior  to the surface is expected to occur at a different pace in accreting sources than in WTTSs.  Moreover, the extended magnetosphere powered by the disk-star magnetic locking, must affect the stellar atmosphere introducing new sources of stirring and turbulence (see e.g. Kivelson \& Russell, 1995). 

\item However, the physical source of the narrow emission component in the He~II profile keeps being uncertain.  It could either be associated with accretion shocks or with atmospheric features. 

\end{itemize}
 
In this section, the possible source of the He~II narrow emission component is analysed as well as some constraints on the extent of the
magnetosphere inferred from the semiforbidden line radiation. 

\subsection{On the source of the narrow component of the He~II line: accretion shocks or bulk atmospheric phenomena?}

The kinematics of the region where the narrow component forms, is clearly distinct from that of the N~V 
or the O~III] lines formation region (see Fig.~8). The dispersion of the narrow emission component of the He~II ranges from 
$\sim 20$~km~s$^{-1}$ to $\sim 60$~km~s$^{-1}$ 
while the dispersion of the N~V line varies varies from $\sim 40$~km~s$^{-1}$ to $\sim 130$~km~s$^{-1}$ for the
same sources (see Fig.~11).  The dispersion of the narrow emission component in the He~II profile has been evaluated as above (see Sect.3) but setting an upper wavelength cut-off 
to reject the contribution of the LDC. Note that even setting this cut-off there is
an unknown contribution from the LDC to the flux.

Moreover, the narrow component seems to be slightly redshifted in {\it all} sources,
from 24~km~s$^{-1}$ in UX~Tau to 37~km~s$^{-1}$ in DE~Tau (see Appendix~1 and Table~7). 
Hence, it would be tempting to suggest that the line is produced in accretion shocks. 
 
Accretion shocks are produced by the impact of the free-falling material from the disk onto the stellar surface.
The kinetic energy of the infalling matter, with typical free-fall speeds of $\sim 300$~km~s$^{-1}$ is
involved into gas heating at the shock front that reaches temperatures of about 1MK. The soft X-ray radiation
from the shock front is expected to photoionize (pre-ionize) the infalling gas column (see i.e. Lamzin 1998, 
G\'omez de Castro \& Lamzin 1999, Muzerolle et al 2001, Orlando et al 2009). Also, the shock front could
back illuminate the stellar surface, becoming a source of atmospheric photoionization to be added to the coronal
X-ray radiation. In this context, the narrow component of the He~II line could be produced very close to the shock front.
The slight, but small redshift, could be interpreted as an indication of the line being formed in postshock material,
and the small width could be caused by thermal broadening; the thermal velocity of fully ionized, solar abundance 
plasma at 50,000~K is 23~km~s$^{-1}$. However, this cannot be concluded from these data alone. Firstly, the apparent
redshift could be caused by the blending of the narrow component with the broad component which is asymmetric
in most sources, {\it e. g.},  with very low or absent flux at bluewards shifted velocities. 
The line broadening is affected by the same problem though no so dramatically given the relative strengths of the 
narrow and broad components. Unfortunately, unless very high S/N profiles ($\sim 100$)
of the semiforbidden and the He~II lines are obtained, any fitting is hampered by these uncertainties. Finally, 
it is also, intriguing that the broadening of the He~II emission line in non-accreting TTSs, such as LkCa~4, LkCa~19
and HBC~427, is comparable to the observed in UX~Tau or in V836~Tau that are TTSs with low accretion rates.
In this respect, it is worth noticing that the correlation between the He~II flux and the accretion luminosity, as derived 
from the U-band excess (Ingleby et al 2009, Gullbring et al 1998) is mild\footnote{Note that the He~II flux
and the accretion luminosity measurements are not simultaneous.  However, the variability of the HeII flux and, in general, of the UV tracers (continuum lines) is typically smaller than a factor of 2 (G\'omez de Castro \& Franqueira 1997; Hu\'elamo et al. 2000).  Measurements of the accretion rate are based in the U-band excess that also varies typically by this amount
(Gullbring et al 1998).}, as shown in Fig.~12 (see also GdCMA2012).

Taking into account all these facts, as well as the good correlation between the He~II and the N~V fluxes (Fig.~9)
and the convergence of the He~II and N~V lines broadenings towards the non-accreting WTTSs, a contribution to the
narrow component of the He~II line from the stellar atmosphere cannot be neglected. In summary, the data analysed in
this work are non-conclusive concerning the source of the narrow component of the He~II line.  For all the reasons mentioned above, 
it is most probable that both physical components, accretion shocks and stellar atmosphere, contribute to the flux.

\subsection{Properties of the LDC}

The profiles produced in the LDC cannot be ascribed to a simple kinematics shared by all sources, neither the radiating plasma 
can be modelled by simple collisional plasma models.  However, some constraints on its  overall physical properties  can be derived from the ratios of the intercombination lines of O~III], Si~III] and C~III]. The plasma density can be constrained from the Si~{\sc iii}]$/$C~{\sc iii}] ratio (G\'omez de Castro \& Verdugo 2001). The detection of O~{\sc iii}]$_{1661,1666}$ can be used to fix up the  temperature since the  O~{\sc iii}]$/$Si~{\sc iii}] ratio is very sensitive to it.

The Si~III] and C~III] lines have been observed with STIS for three sources in the sample: RW~Aur (G\'omez de Castro \& Verdugo 2003, hereafter GdCV2003), DE~Tau and T~Tau (G\'omez de Castro et al 2003). T~Tau profiles display a non-negligible contribution from the large scale jet. DE~Tau Si~III] and C~III] profiles are rather narrow and similar to the O~III] line. 
From RW~Aur study, GdCV2003  pointed out that the emitting volume is clumpy with a rather small filling factor, as otherwise expected if magnetospheric radiation is produced in plasma filaments and clumps. 
This clumpy nature together with the broad temperature range covered by the various spectral tracers suggest that the excitation mechanism could be photoionization instead of collisional excitation.

GdCV2003 produced two grids of photonionization models to explore possible regimes for line excitation making
use of CLOUDY (Ferland 1996), a code designed to simulate emission line regions in astrophysical environments.
The first set assumed that LDC had a belt like geometry, being illuminated by the ambient X-ray radiation field:
soft and hard components, at $3.5\times 10^6$~K and 2.8$\times 10^7$~K respectively, with a total X-rays luminosity of 
$3\times10^{29}$~erg~s$^{-1}$. This model turned out to be unable to reproduce comparable strengths of the three
spectral tracers. However, if the O~III], Si~III] and C~III] emission is assumed to be produced in dense gas 
around small X-ray sources, such as reconnecting loops, the lines ratios could be reproduced for electron densities 
of $n_e \geq 10^{11}$~cm$^{-3}$. For soft X-ray sources, with T$_e = 10^6$~K, luminosities of $10^{27}$~erg~s$^{-1}$ and
radii of 10$^8$~cm$^{-3}$, the inferred O~III] emissivity is $\sim 10^{-3.5}$ erg~s~cm$^{-2}$. Using this one as a fiducial 
value, an estimate of the LDC volume, $V_{LDC}$ could be derived from the line strength as, 
$\frac {V_{LDC}} {\eta} = \frac {F(O III]) 4 \pi d^2}{\epsilon _{O III]}}$ where $F(O III])$ is the reddening corrected line flux and $\eta$ is the filling factor of the hot plasma.  For filling factors of 10, and assuming that the emission
is concentrated in a spherical shell of radius R$_{LDC}$ and thickness 0.01R$_{LDC}$, LDC radii from 4 to 9 R$_{\odot}$
are inferred. These values are within a  factor of 1.5 of the magnetospheric radii derived
from accretion luminosities for stars with known magnetic fields (Johns-Krull 2007, GdCMA2012). Unfortunately, the uncertainties in the plasma distribution prevent more detailed evaluations.

\section{Conclusions}

The UV radiation from N~V, He~II and O~III] is dominated by the contribution of three main components of the 
TTSs atmospheric/magnetospheric environment: the magnetospheric flow of infalling matter, the disturbed upper
atmospheric layers and the accretion shock.  

The diffuse magnetospheric plasma (LDC) is best traced by the O~III] line: it produces asymmetric profiles, 
preferentially red-shifted. He~II emission is observed from the same kinematical structures that radiate in O~III]. 
This plasma is not excited by thermal collisions at a single electron temperature. In fact, there are 
indications that photoionization processes could be significant.

The hot, dense layers of the stellar atmosphere are best traced by the N~V line. However, the line dispersion 
increases steadily with the accretion rate suggesting a connection between the disturbances in the upper 
atmospheric layers and the accretion flow. 

The He~II flux is dominated by a narrow emission component of uncertain origin. In this work, we have presented
evidences indicating that it may be formed in  hot postshock material in accretion shocks but also, evidences 
of its connection with atmospheric tracers. Both accretion shocks and the upper atmospheric layers seem to 
contribute to the narrow component of the He~II profile. 

The anticorrelation between X-ray and UV flux found but GdCMA2012 has been confirmed, suggesting that
the dissipation of magnetic energy proceeds in the TTSs differently than in main 
sequence stars.  The denser environment produced by mass accretion (see i.e. Petrov et al. 2011)
seems to favour the ultraviolet channel for the dissipation of the magnetic energy excess.

All the observations indicate that the UV radiation field during PMS evolution is much harder than the 
usually implemented in the modelling of protostellar disks evolution. An example of its effect in the 
dust grains charging and charging profile can be found in Pedersen \& G\'omez de Castro 2011. Theoretical 
modelling of protostellar disks chemistry and life generation environments should take this fact into account.

\acknowledgments

Kevin France brought to my attention the HST/COS data corresponding to AA~Tau. 
The analysis of the data pointed out that the 1640\AA\ jump in the HST/ACS data
was dominated by an unexpectedly strong He~II line. This article has grown from 
extending this analysis to the TTSs observed with HST/COS. I would like to thank
an anonymous referee for suggesting the use of the H$_2$ lines to set the
zero of the wavelength scale. This work has been 
partly funded by the Ministerio de Economia y Competitividad of Spain through grant 
AYA2011-29754-C03-C01.

\appendix
\section{The velocity of the He~II narrow emission component}

The narrow emission component of the He~II line seems to be systematically red-shifted with respect to the H$_2$ lines. The shifts are small ($\sim 0.16$~\AA\ or 29~km~s$^{-1}$) as shown in Table~8. The zero of the wavelength scale has been set with the (R(3)~1-7) 1489.636~\AA\ and (P(5)~1-7) 1504.845~\AA\  H$_2$ lines because they were observed in most of the stars. These lines are far from the 1640~\AA\ He~II line and, in principle, small uncertainties in the wavelength calibration could drive to these shifts. However, as shown in Figures 14 and 15, this is not the case. The P(17)3-9 line is plotted for the whole sample in Figure~14 though only is clearly observed in: DM~Tau, UX~Tau~A, AA~Tau, GM~Aur and DR~Tau (the H$_2$ emission from RW~Aur cannot be used for this purpose). In all cases, the H$_2$ transition is at rest. In Figure~15, the C~I[uv1] multiplet is plotted. It is observed as narrow emission lines in DN~Tau, DM~Tau, UX~Tau~A and DF~Tau. In all cases,
the lines are at rest wavelength. This red-shift is observed both in CTTSs and intermediate 
objects (it cannot be measured in WTTSs because of the lack of H$_2$ emission). 
Thus, unless the P(14)~3-10 H$_2$ transition, that it is blended with the He~II line, is unusually strong, it must be concluded that the narrow emission component of the HeII line is red-shifted in most sources. Note however, that the blending with
the broad component could produce an apparent redshift if this broad component is asymmetric.

{\it Facilities:} \facility{HST (COS)}, \facility{HST (STIS)}, \facility{XMM-Newton (EPIC)}.


\newpage

\begin{figure}[h]
   \includegraphics[width=1\textwidth]{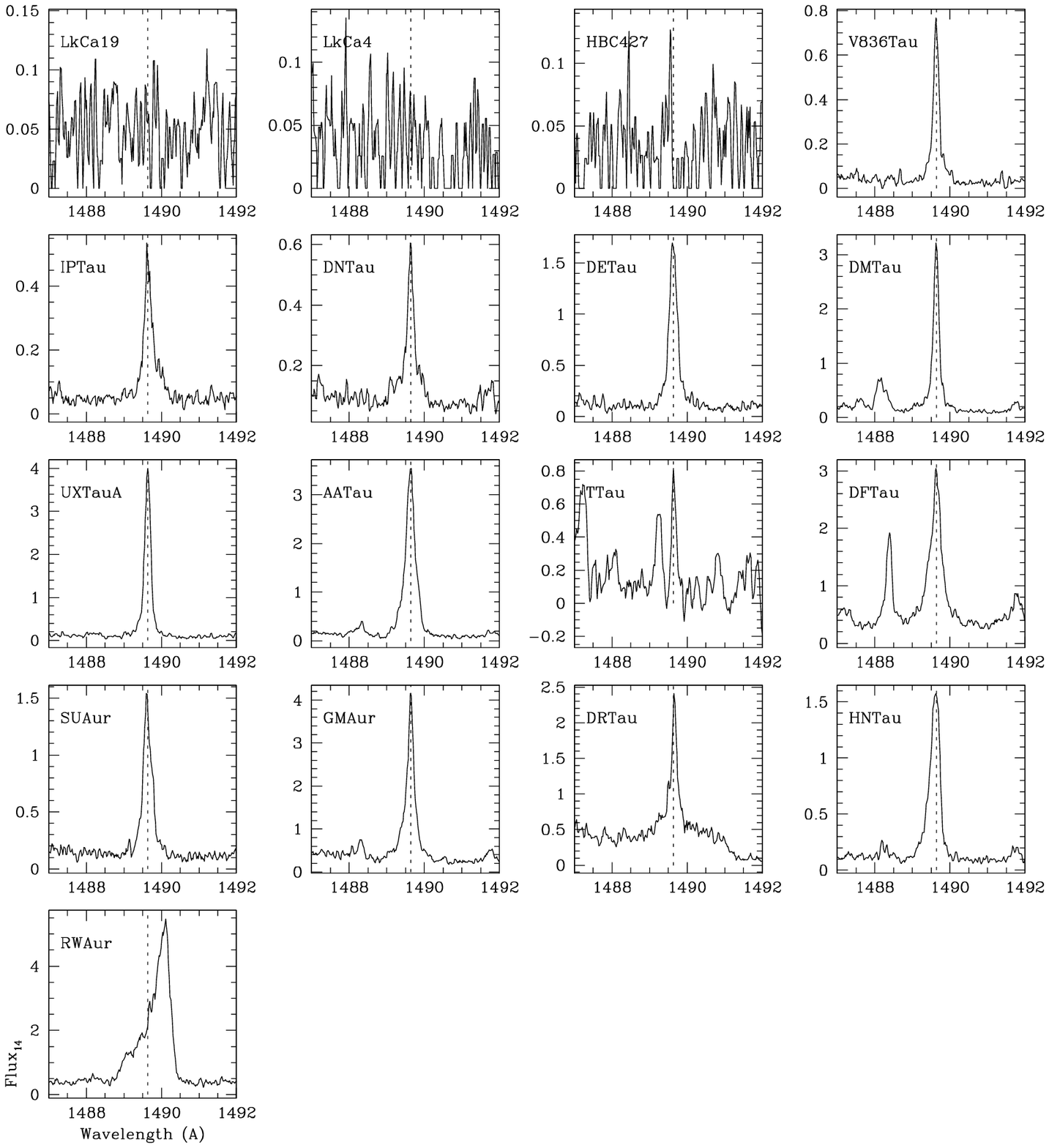}
 \caption{Profiles of the (R(3)~1-7) 1489.636~\AA\ H$_2$ line of the TTSs studied in this work. The zero of the wavelength scale has been set using H$_2$ lines as reference (see text). Notice the asymmetry and broadening of RW~Aur profile that prevents its use for this purpose.  Fluxes are given in units of 10$^{-14}$~erg~s$^{-1}$~cm$^{-2}$~\AA $^{-1}$ (F$_{14}$)}.
\label{fig:he2}
\end{figure}

\newpage

\begin{figure}[h]
   \includegraphics[width=1\textwidth]{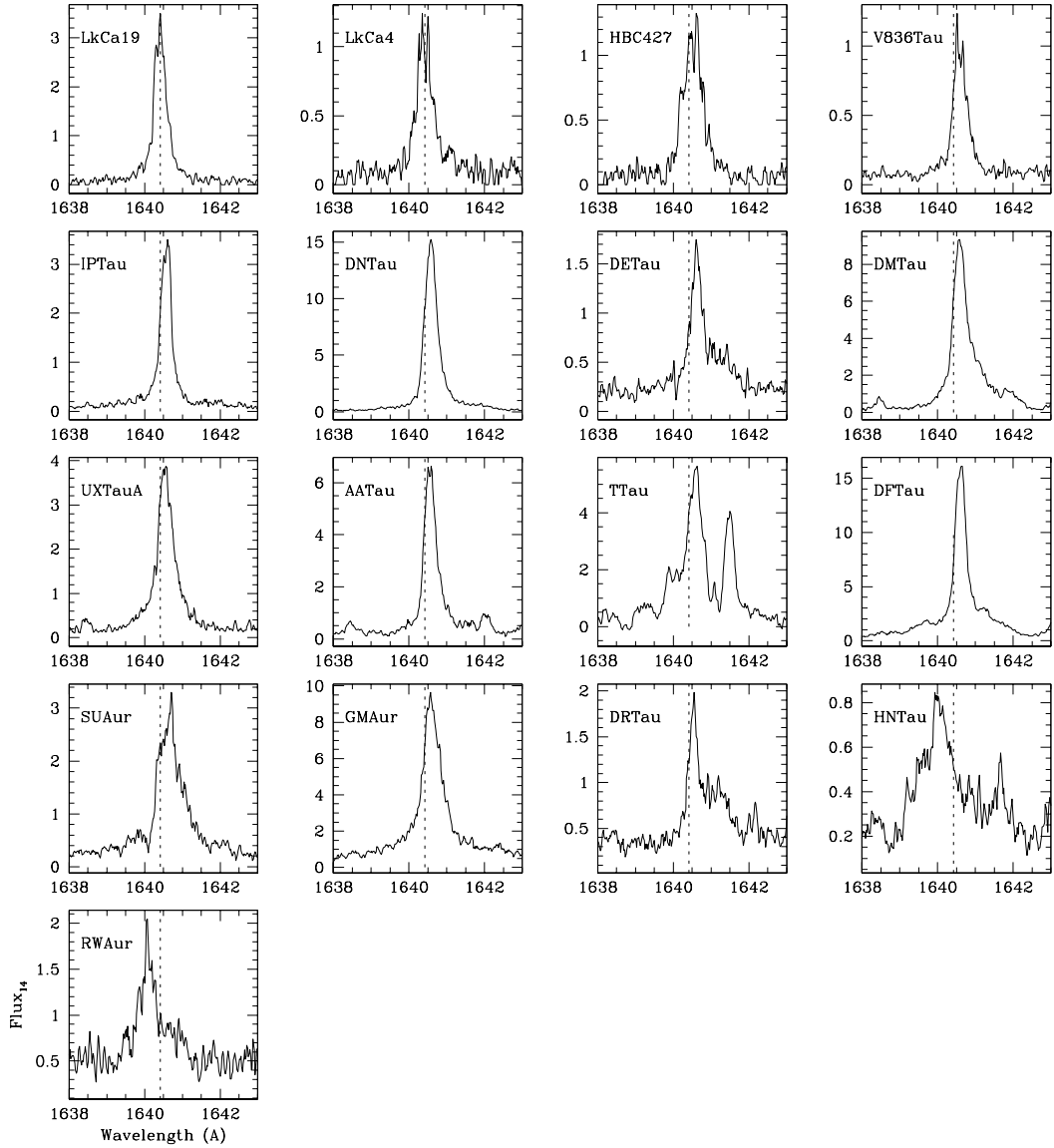}
 \caption{He~II profiles of the TTSs in Taurus observed with HST.
The rest wavelength of the He~II line is marked for reference (see also Figure~1). Fluxes are given in units of 10$^{-14}$~erg~s$^{-1}$~cm$^{-2}$~\AA $^{-1}$ (F$_{14}$).}
\label{fig:h2_1}
\end{figure}

\begin{figure}[h]
   \includegraphics[width=0.8\textwidth]{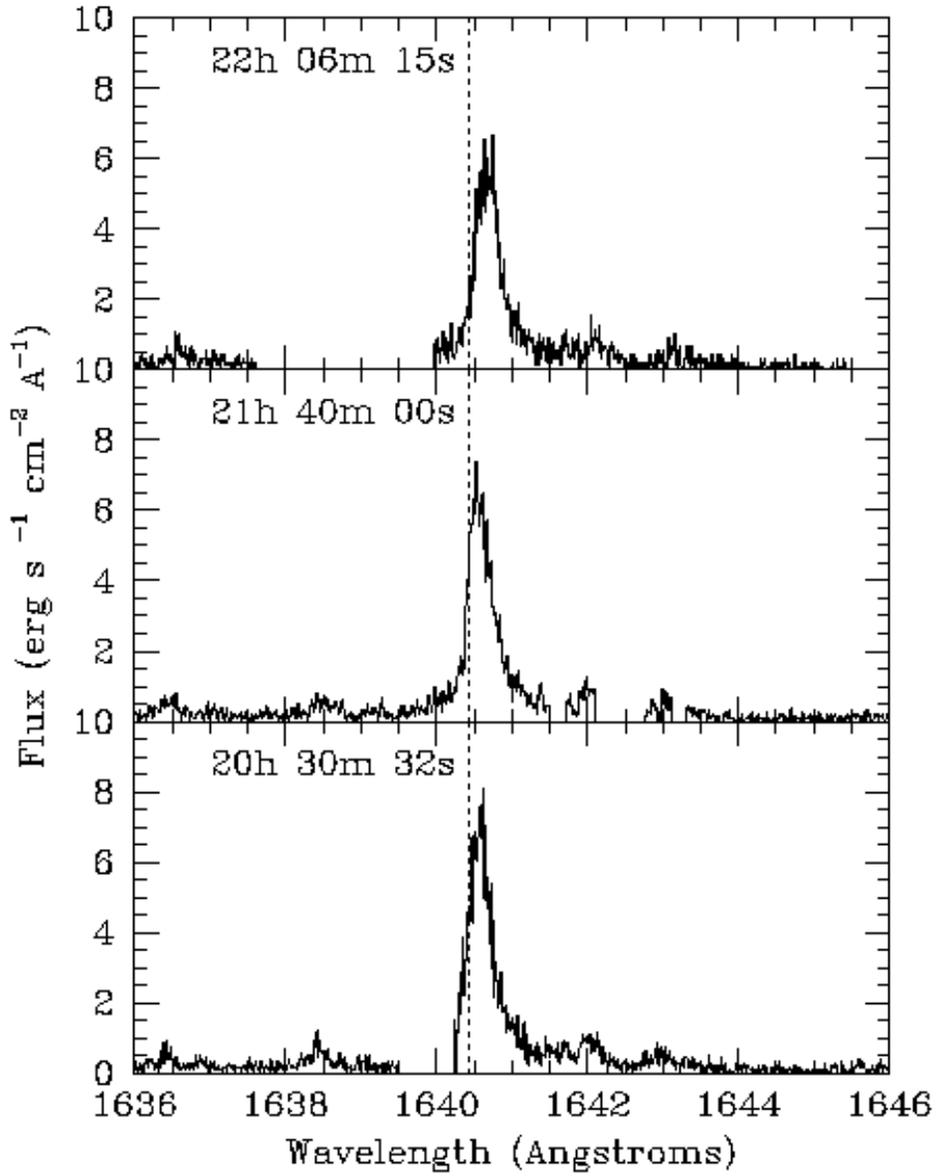}
 \caption{The three observations of the He~II line in AA~Tau with HST/COS. The rest wavelength of the He~II line is marked for reference. During the observations the guide star acquisition
failed. Note the broadening and shift of the central peak from the first to the last observation.}
\label{fig:aatau}
\end{figure}

\newpage

\begin{figure}[h]
   \includegraphics[width=1\textwidth]{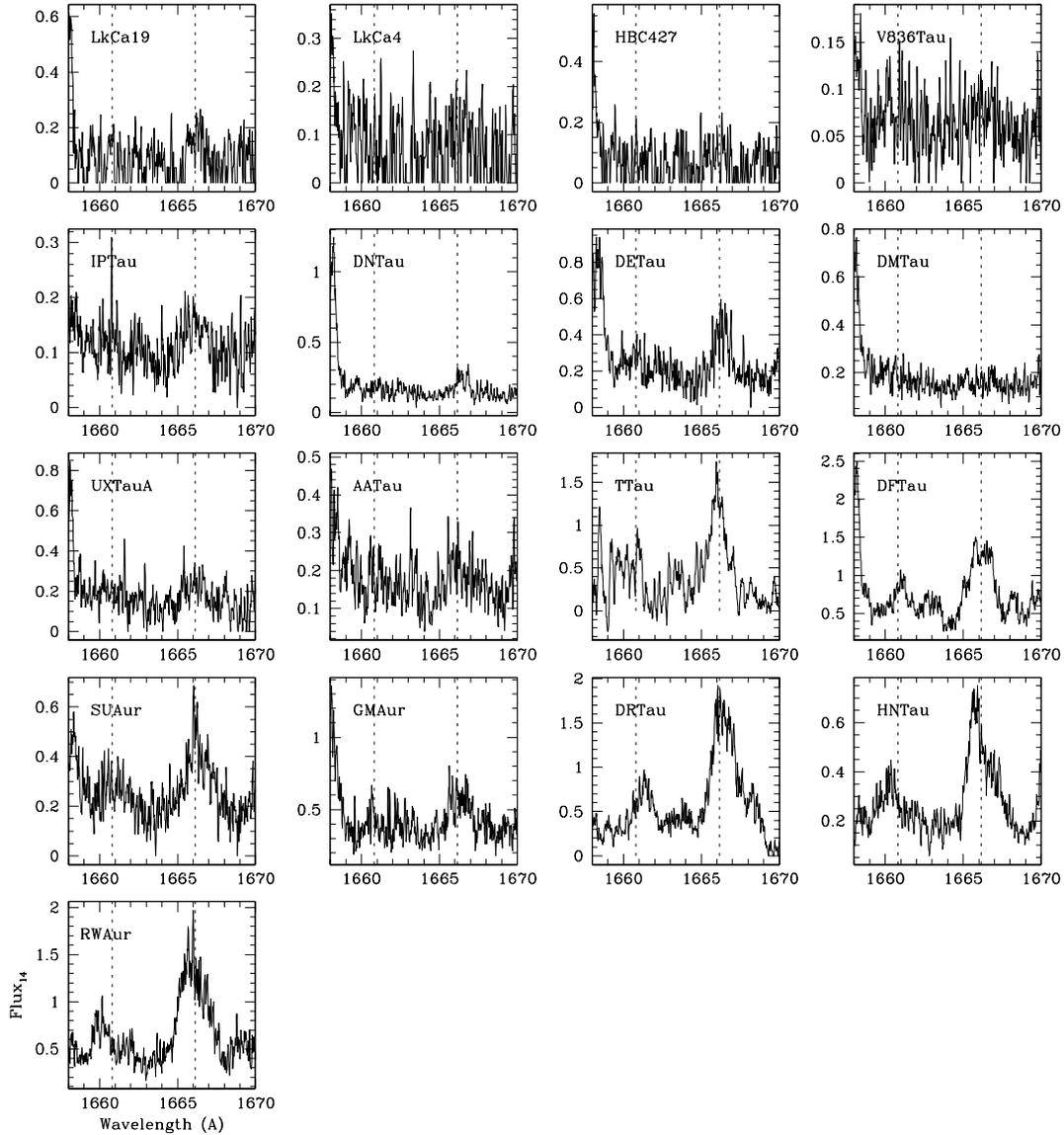}
 \caption{O~III] profiles of the TTSs in Taurus observed with HST.
The rest wavelength of the two lines in the multiplet is marked for reference
(see also Table~1). Fluxes are given in units of 10$^{-14}$~erg~s$^{-1}$~cm$^{-2}$~\AA $^{-1}$ (F$_{14}$).}
\label{fig:o3}
\end{figure}

\newpage

\begin{figure}[h]
   \includegraphics[width=1\textwidth]{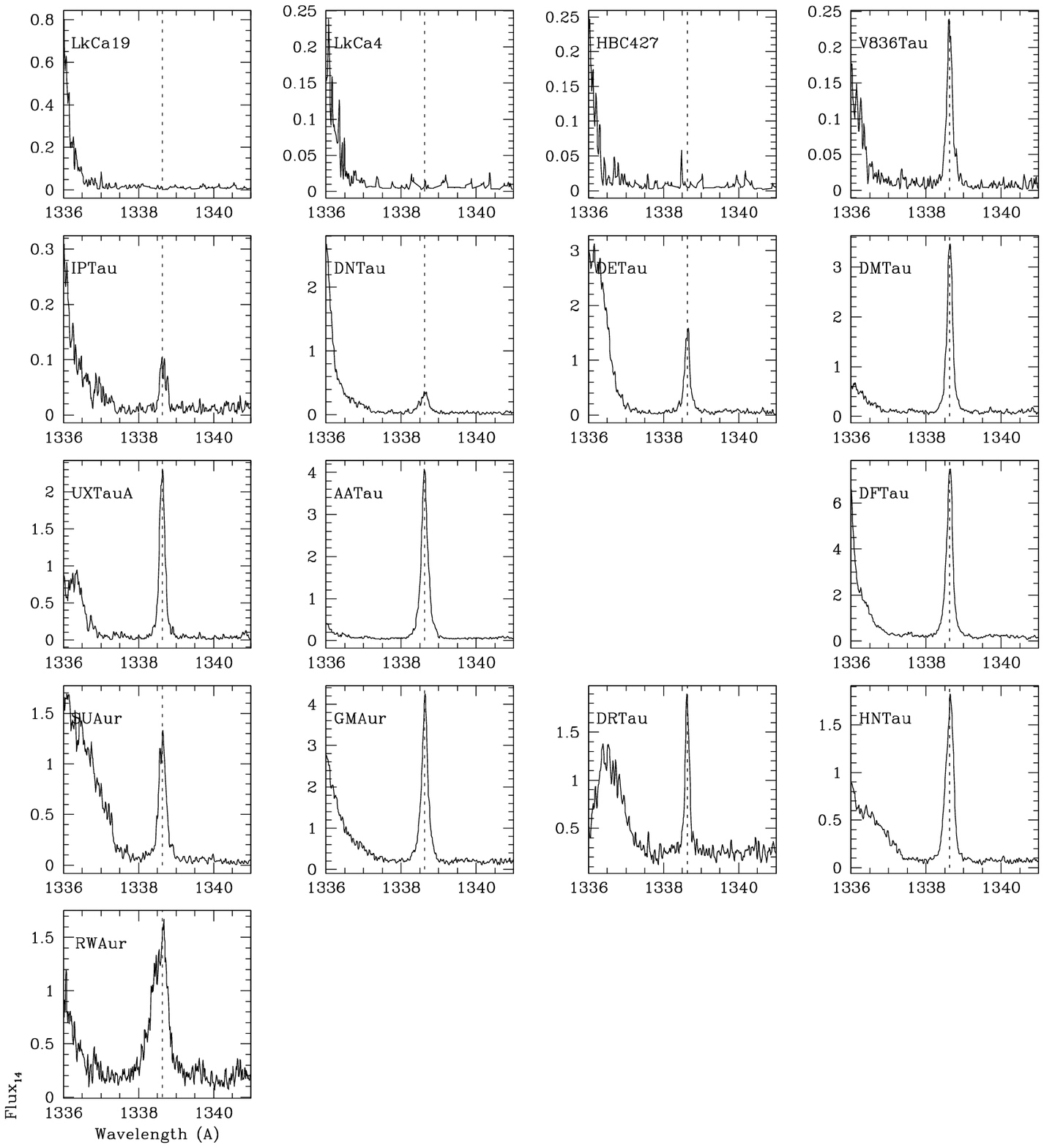}
 \caption{Profiles of the (P(2)~0-4) 1338.63~\AA\  H$_2$ line of the TTSs studied in this work. The zero of the wavelength scale has been
set using H$_2$ lines as reference (see text). Notice the asymmetry and broadening of the RW~Aur profile that prevents its use for
this purpose. Fluxes are given in units of 10$^{-14}$~erg~s$^{-1}$~cm$^{-2}$~\AA $^{-1}$ (F$_{14}$).}
\label{fig:h2_2}
\end{figure}

\newpage

\begin{figure}[h]
   \includegraphics[width=1\textwidth]{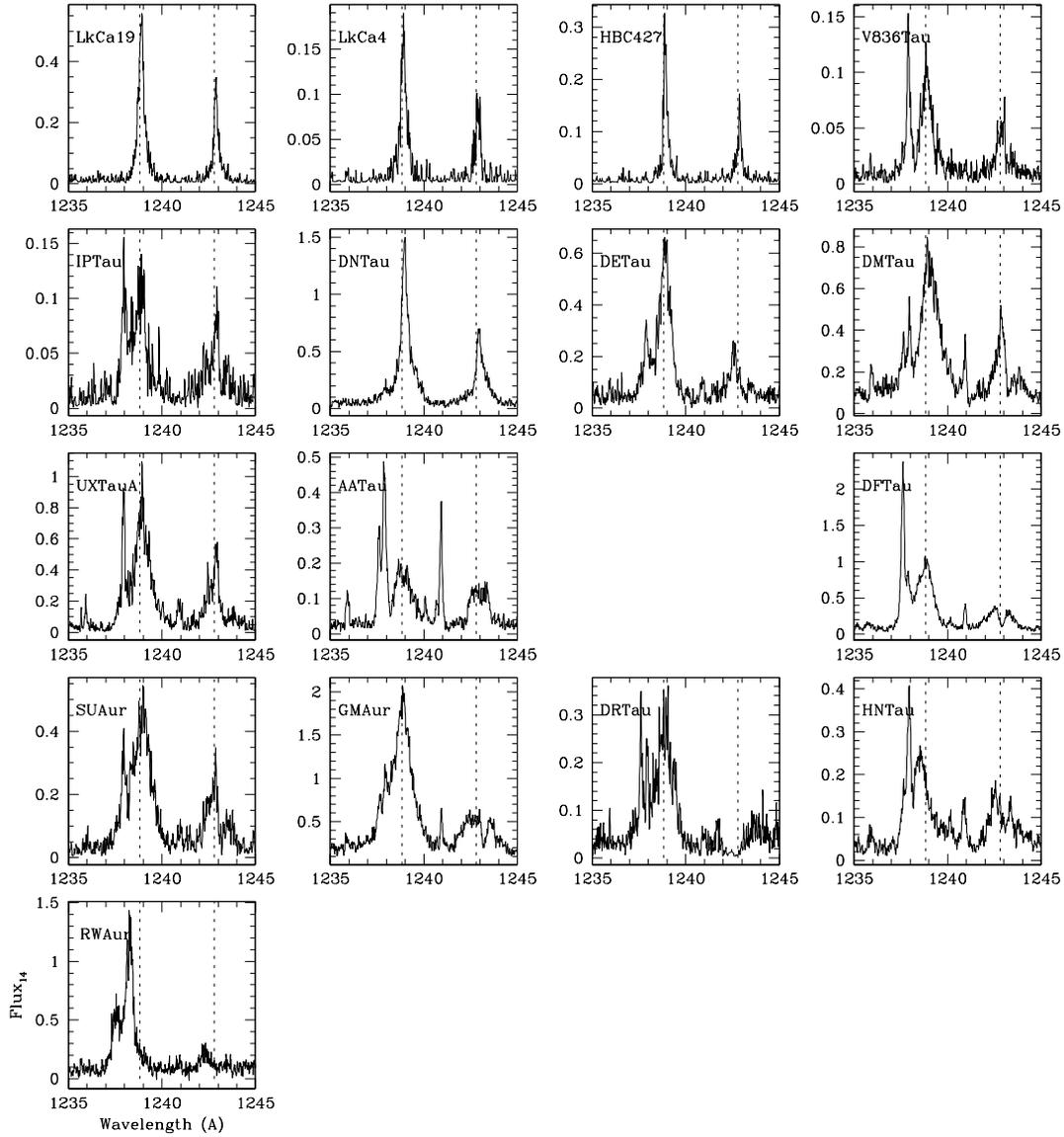}
 \caption{N~V profiles of the TTSs in Taurus. The rest wavelengths of the N~V lines are marked for reference (see also Figure~5). Fluxes are given in units of 10$^{-14}$~erg~s$^{-1}$~cm$^{-2}$~\AA $^{-1}$ (F$_{14}$). }
\label{fig:n5}
\end{figure}

\newpage

\begin{figure}[h]
   \includegraphics[width=0.80\textwidth]{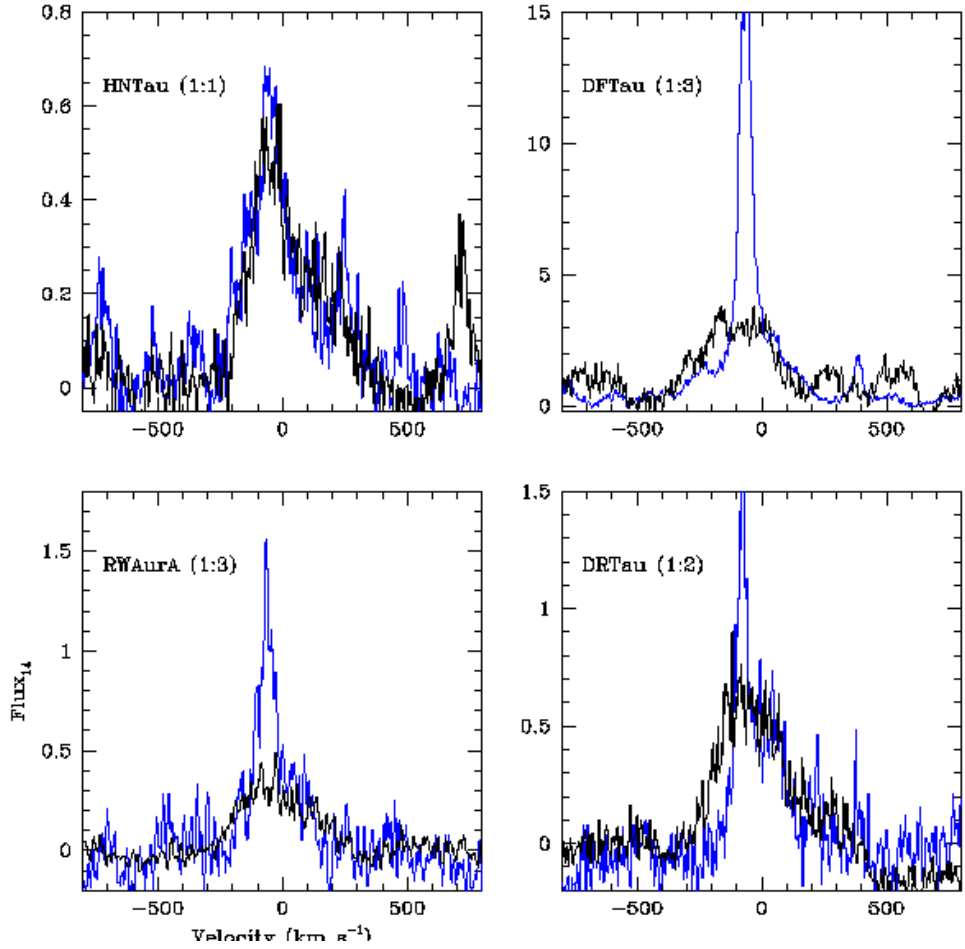}
 \caption{High SNR O~III] profiles (black)  over-plotted on the respective He~II profiles (blue). The O~III] profiles have been re-scaled for comparison. The scaling factor is  indicated in the figure for every profile.}
\label{fig:comp}
\end{figure}

\newpage

\begin{figure}[h]
   \includegraphics[width=0.70\textwidth]{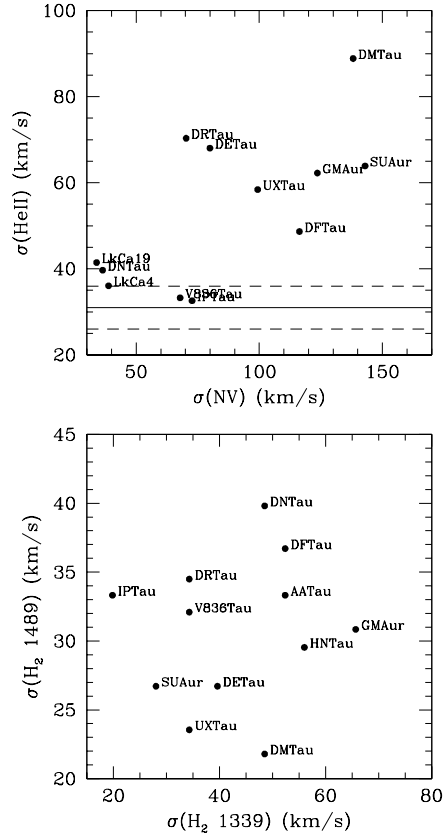}
 \caption{{\it Top panel}: Broadening of the N~V line (dispersion) compared with the broadening of the He~II line. The broadening is defined as the dispersion of the profile (see text). For comparison the broadening of the H$_2$ 1489~\AA\ and 1389~\AA\ lines are plotted in the bottom panel. The dispersion of the N~V line is larger 
than the measured in the H$_2$  1389~\AA\ line in all sources. However the dispersion of  H$_2$ 1489~\AA\ is comparable to the observed in the He~II line in WTTSs and transitional objects. The average $\sigma$(H$_2$(1489)) = 31$\pm$5 km~s$^{-1}$ is represented in the top panel; the solid line represents the average and the error band is marked with dashed lines. Circled
sources have a significant contribution from the LDC to the He~II flux.}
\label{fig:sigman5}
\end{figure}

\newpage
\begin{figure}[h]
   \includegraphics[width=0.80\textwidth]{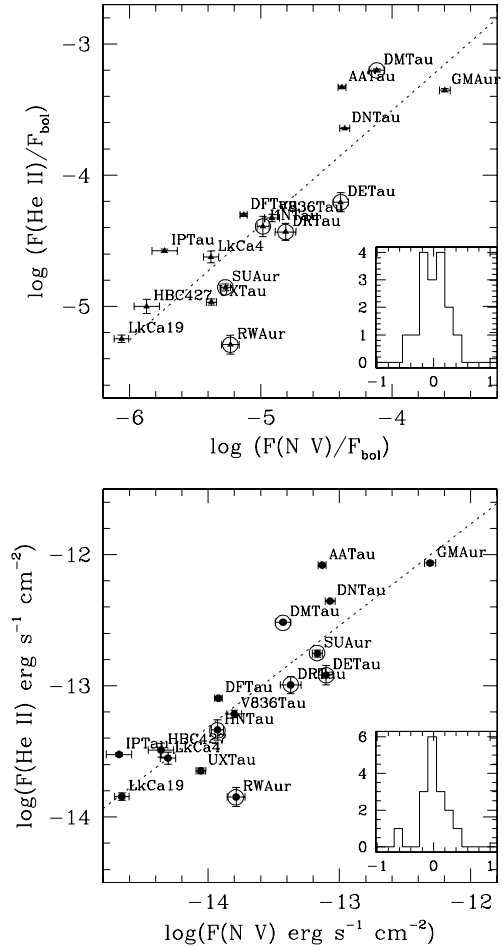}
 \caption{{\it Bottom}: the He~II flux is plotted versus the N~V flux. {\it Top}: the He~II normalised flux 
is plotted against the N~V normalised flux. Fluxes are extinction corrected. The small insets at the 
bottom-right corner are the histograms representing the distance of the sources to the main regression line, 
plotted as a dashed line. Circled
sources have a significant contribution from the LDC to the He~II flux.}
\label{fig:n5he}
\end{figure}

\newpage

\begin{figure}[h]
   \includegraphics[width=0.40\textwidth]{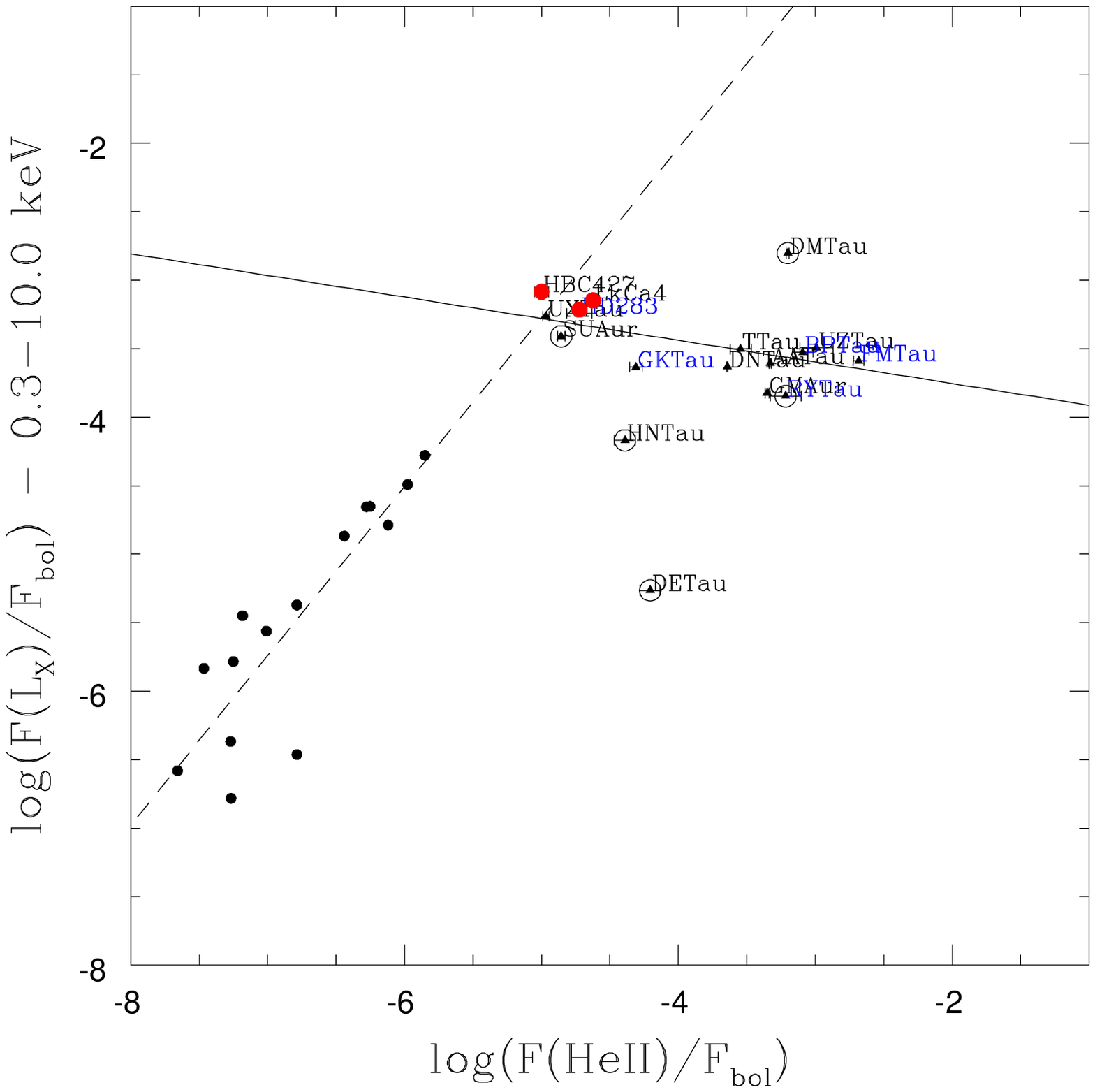}\\
   \includegraphics[width=0.40\textwidth]{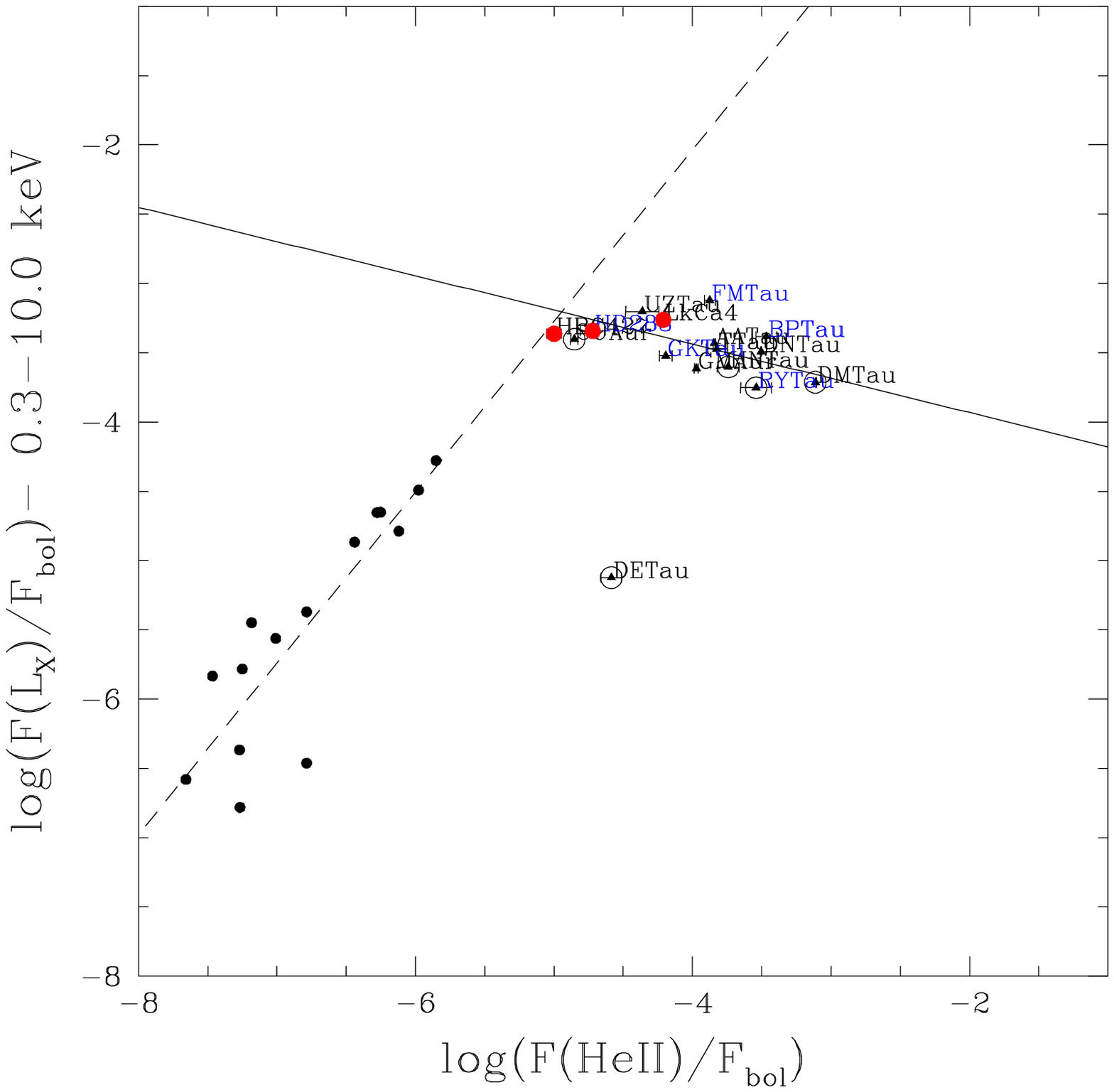}
 \caption{Normalised X-ray flux versus normalised He~II flux for TTSs compared with main sequence stars cool stars (MSCSs). TTSs are plotted with triangles and the error bars are marked. MSCSs are plotted with filled circles (data from Ayres et al 1995 and Linsky et al 1982, as described in GdCMA2012). The MSCSs regression line is plotted with a dashed line and the TTSs regression line with a continuous line.
The stars names are indicated in black for F(He~II) measurements based on high resolution data (this work) and in blue for GdCMA2012 measurements based on low dispersion data.  Red dots mark WTTSs location. {\it Top:} A$_V$, f$_{bol}$ and L$_X$ as in Table~3. Circled
sources have a significant contribution from the LDC to the He~II flux.
{\it Bottom:} A$_V$, f$_{bol}$ and L$_X$ as in Table~2 of Hao Yang et al. 2012. }
\label{fig:ff}
\end{figure}

\newpage
\begin{figure}[h]
\includegraphics[width=1.0\textwidth]{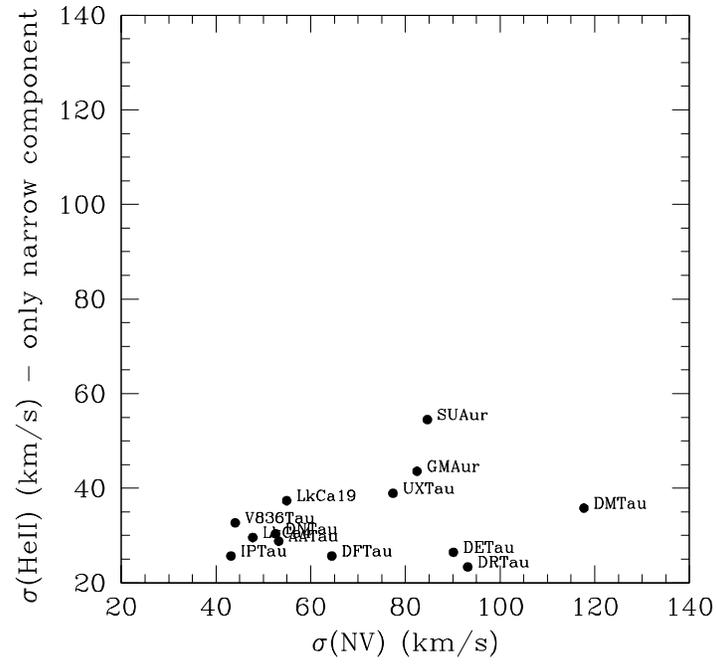}
\caption{As Fig.~8 but the dispersion of the He~II line is evaluated only for the narrow component of the profile.}
\label{tracers}
\end{figure}

\newpage
\begin{figure}[h]
\includegraphics[width=0.55\textwidth]{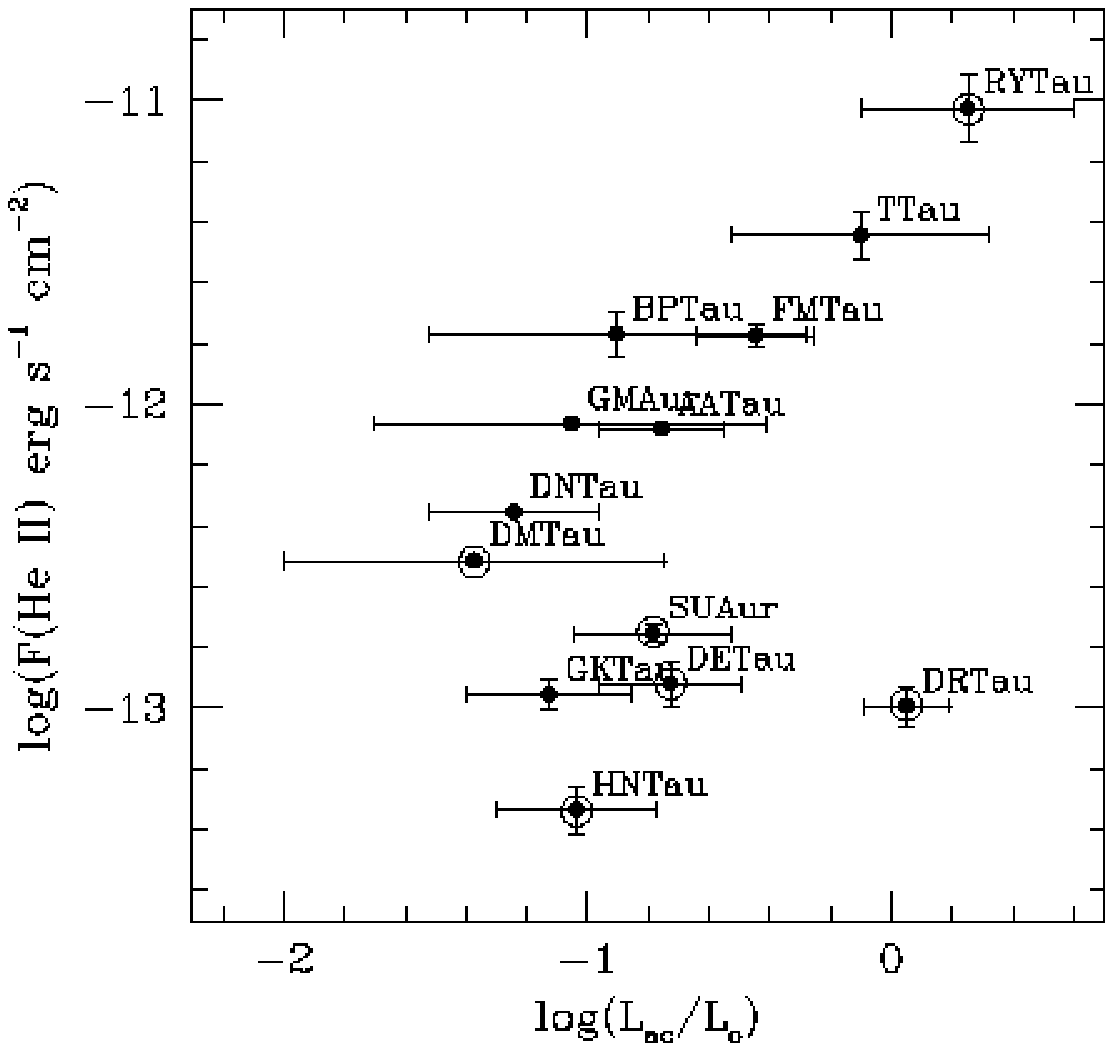}\\
\includegraphics[width=0.55\textwidth]{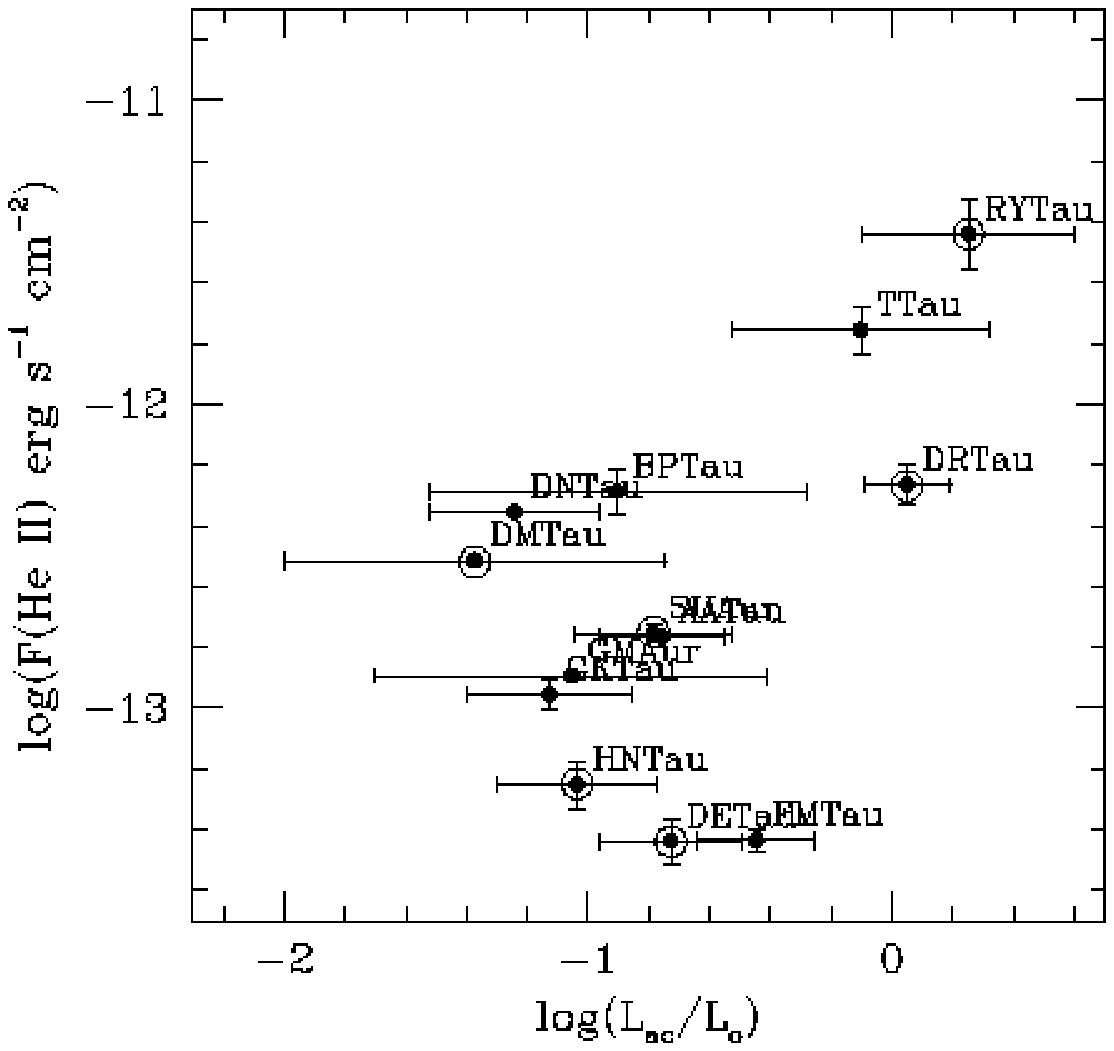}
\caption{{\it Top panel:} He~II fluxes versus accretion luminosities using extinctions from Table~3. {\it Bottom panel:}
same as the top panel but with extinctions from Yang et al (2012). Note that accretion luminosities have not been measured
simultaneously; typical variations can account for a factor as large as $\sim$2 (0.3 in logarithmic scale).}
\label{he2acretion}
\end{figure}

\newpage

\begin{figure}[h]
   \includegraphics[width=1.0\textwidth]{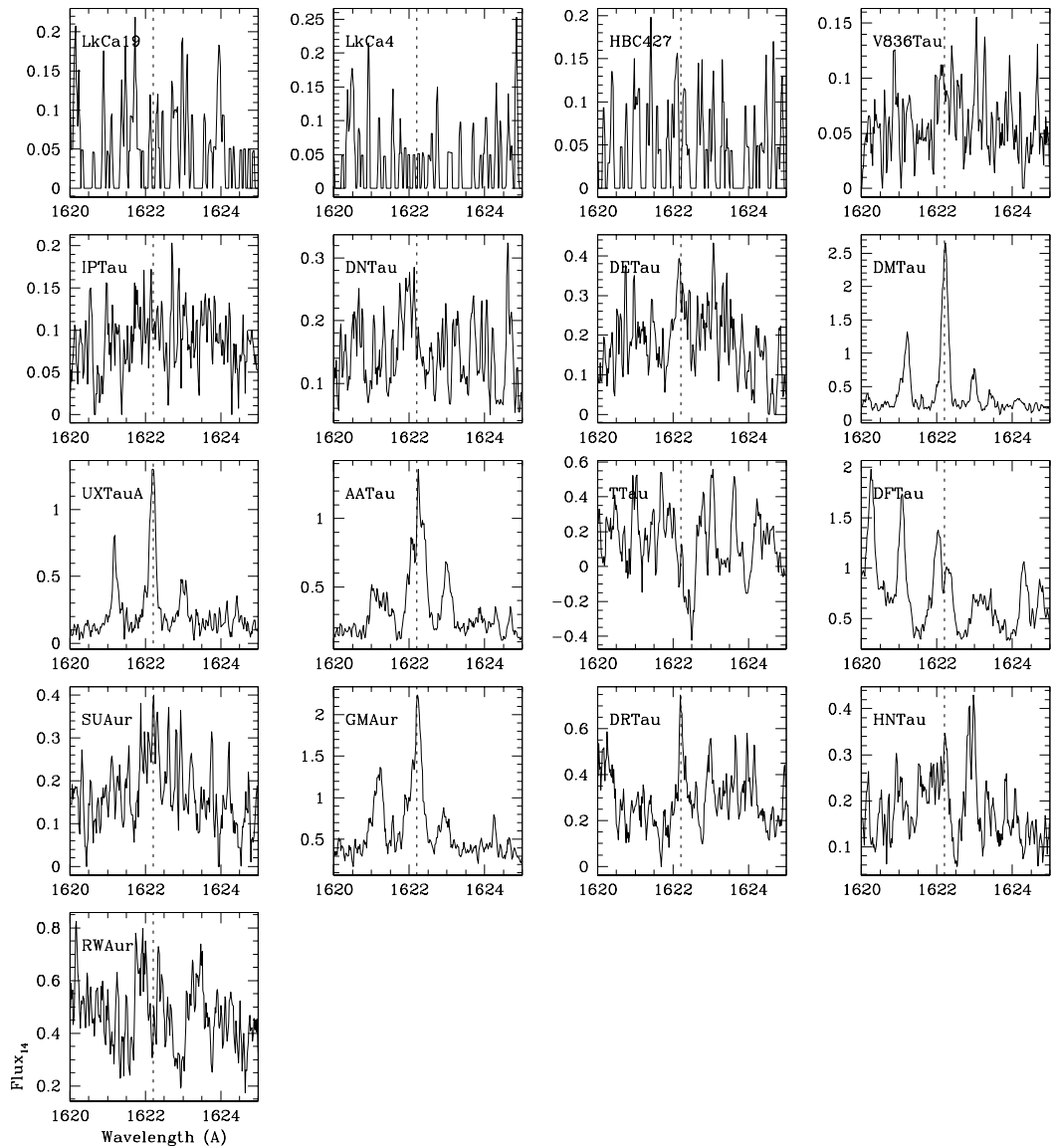}
 \caption{P(17)3-9 H$_2$ profiles of the TTSs in Taurus observed with HST.
The rest wavelength is marked for reference. Fluxes are given in units of 10$^{-14}$~erg~s$^{-1}$~cm$^{-2}$~\AA $^{-1}$ (F$_{14}$).}
\label{fig:h2_3}
\end{figure}

\begin{figure}[h]
   \includegraphics[width=1.0\textwidth]{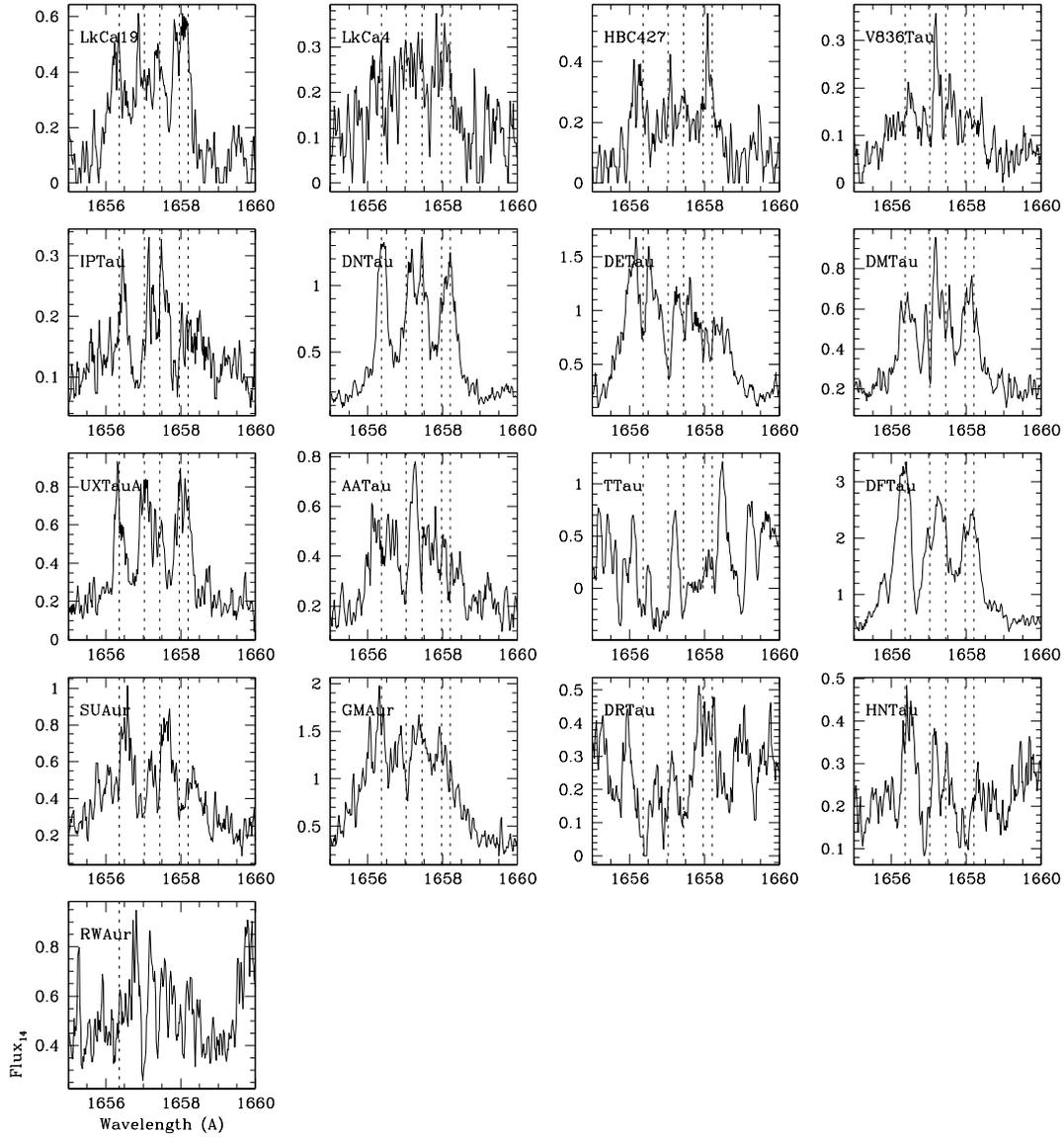}
 \caption{C~I [uv1] profiles of the TTSs in Taurus observed with HST.
The rest wavelength of the lines in the multiplet is marked for reference. Fluxes are given in units of 10$^{-14}$~erg~s$^{-1}$~cm$^{-2}$~\AA $^{-1}$ (F$_{14}$).}
\label{fig:c1}
\end{figure}


\newpage
\topmargin=-40pt
\voffset=-20pt
\begin{table*}
\begin{tiny}
\begin{center}
\caption{HST/COS observations of the He~II line.\label{tab:table1}}
\begin{tabular}{lllllll}
\hline\hline
 Star & Instrument/ & Observation  &  Start Time          & Exposure   & Spec. Initial & Spec. Final  \\
      & Grating     & ID           & (yyyy-mm-dd hh:mm:ss)& Time (sec) & Wavelength (\AA )& Wavelength (\AA )\\
\hline
SU~Aur$^3$ & COS/G160M & LB6B11010 &2011-03-25 08:01:38 & 622.144 & 1387.748 & 1748.301  \\
	& COS/G160M & LB6B11020 &2011-03-25 08:15:08 & 622.144 & 1410.706 & 1771.310  \\
	& COS/G160M & LB6B11030 &2011-03-25 09:06:57 & 515.008 & 1434.646 & 1795.304  \\
T~Tau$^1$ & STIS/E140M & O5E304020 & 2000-09-08 09:11:01 	& 2630.173 & 1140.000 & 1735.000 \\
	& STIS/E140M & O5E304040 & 2000-09-08 10:56:04 	& 2320.191 & 1140.000 &	1735.000 \\
	& STIS/E140M & O5E304050 & 2000-09-08 12:23:56 	& 2630.185 & 1140.000 & 1735.000 \\
	& STIS/E140M & O5E304060 & 2000-09-08 14:00:24 	& 2630.164 & 1140.000 & 1735.000 \\
LkCa~19$^3$ & COS/G160M & LB6B28030 &2011-03-31 01:21:43 & 630.112 & 1387.740 & 1748.270 \\ 
 	& COS/G160M & LB6B28040 &2011-03-31 02:33:25 & 630.112 & 1410.699 & 1771.292 \\
 	& COS/G160M & LB6B28050 &2011-03-31 02:47:03 & 630.112 & 1434.577 & 1795.224 \\
GM~Aur$^3$ 	& COS/G160M & LB6B01030 &2010-08-19 16:53:54 & 620.192 & 1387.921 & 1748.541 \\
	& COS/G160M & LB6B01040 &2010-08-19 17:07:22 & 620.192 & 1410.957 & 1771.592 \\ 
	& COS/G160M & LB6B01050 &2010-08-19 17:20:50 & 620.192 & 1431.254 & 1791.979 \\ 
RW~Aur~A$^3$& COS/G160M & LB6B15010 &2011-03-25 02:51:01 & 539.008 & 1390.596 & 1751.126 \\
 	& COS/G160M & LB6B15020 &2011-03-25 03:03:08 & 538.976 & 1413.616 & 1774.208 \\
	& COS/G160M & LB6B15030 &2011-03-25 03:15:43 & 538.976 & 1431.426 & 1792.083 \\
HN~Tau$^3$	& COS/G160M & LB6B09010 &2010-02-10 12:49:14 &1736.352 & 1388.187 & 1755.142 \\
	& COS/G160M & LB6B09020 &2010-02-10 14:11:51 &1396.160 & 1411.659 & 1772.251 \\
	& COS/G160M & LB6B09030 &2010-02-10 15:32:12 &1396.160 & 1435.611 & 1796.256 \\
UX~Tau~A$^3$& COS/G160M & LB6B13030 &2010-12-12 14:44:10 & 647.168 & 1387.800 & 1748.355 \\
	& COS/G160M & LB6B13040 &2010-12-12 14:58:03 & 647.104 & 1410.723 & 1771.317 \\
	& COS/G160M & LB6B13050 &2010-12-12 15:11:58 & 647.200 & 1434.591 & 1795.276 \\
HBC~427$^3$ & COS/G160M & LB6B26030 &2011-03-30 02:34:43 & 644.992 & 1387.800 & 1748.355 \\
	& COS/G160M & LB6B26040 &2011-03-30 02:48:36 & 645.120 & 1410.881 & 1771.450 \\
	& COS/G160M & LB6B26050 &2011-03-30 03:02:29 & 645.088 & 1434.662 & 1795.309 \\
AA~Tau$^3$  & COS/G160M & LB6B07010 &2011-01-06 20:30:32 & 1418.176& 1387.834 & 1748.363 \\ 
	& COS/G160M & LB6B07020 &2011-01-06 21:40:00 & 1387.168& 1410.854 & 1771.435 \\ 
	& COS/G160M & LB6B07030 &2011-01-06 22:06:15 & 1387.136& 1434.611 & 1795.257 \\ 
DR~Tau$^3$  & COS/G160M & LB6B14010 &2010-02-15 11:08:53 & 582.016 & 1388.651 & 1749.265 \\
	& COS/G160M & LB6B14020 &2010-02-15 12:04:39 & 581.952 & 1411.842 & 1772.470 \\
	& COS/G160M & LB6B14030 &2010-02-15 12:17:29 & 581.984 & 1435.647 & 1796.340 \\
LkCa~4$^3$  & COS/G160M & LB6B27030 &2011-03-30 07:32:26 & 600.192 & 1387.798 & 1748.330 \\
        & COS/G160M & LB6B27040 &2011-03-30 07:45:34 & 600.160 & 1410.781 & 1771.352 \\
        & COS/G160M & LB6B27050 &2011-03-30 07:58:42 & 600.128 & 1434.635 & 1795.284 \\
V836~Tau$^3$& COS/G160M & LB6B06010 &2011-02-05 05:06:58 & 1480.192& 1393.645 & 1754.155 \\
	& COS/G160M & LB6B06020 &2011-02-05 06:12:31 & 1391.136& 1410.840 & 1771.423 \\
	& COS/G160M & LB6B06030 &2011-02-05 06:38:50 & 1391.200& 1434.670 & 1795.307 \\
DE~Tau$^3$ 	& COS/G160M & LB6B08030 &2010-08-20 15:14:30 & 617.152 & 1387.865 & 1748.487 \\ 
	& COS/G160M & LB6B08040 &2010-08-20 15:27:55 & 617.184 & 1410.962 & 1771.598 \\
	& COS/G160M & LB6B08050 &2010-08-20 15:41:20 & 617.088 & 1434.735 & 1795.449 \\
IP~Tau$^3$ 	& COS/G160M & LB6B05010 &2011-03-21 08:11:35 & 1179.136& 1393.598 & 1754.105 \\
	& COS/G160M & LB6B05020 &2011-03-21 09:15:35 & 972.192 & 1407.600 & 1768.179 \\
	& COS/G160M & LB6B05030 &2011-03-21 09:34:45 & 972.128 & 1440.509 & 1801.117 \\
DF Tau$^2$  & COS/G160M & LB3Q020A0 &2010-01-11 13:07:29 & 1324.192& 1435.545 & 1796.229 \\
	& COS/G160M & LB3Q02090 &2010-01-11 12:00:02 & 1408.192& 1423.424 & 1784.071 \\
	& COS/G160M & LB3Q02080 &2010-01-11 11:32:10 & 1408.192& 1411.793 & 1772.421 \\
	& COS/G160M & LB3Q02070 &2010-01-11 10:24:56 & 1408.192& 1400.406 & 1761.021 \\
DM~Tau$^3$	& COS/G160M & LB6B02030 &2010-08-22 17:23:51 & 978.176 & 1387.890 & 1748.500 \\
	& COS/G160M & LB6B02040 &2010-08-22 18:26:31 & 1396.128& 1410.865 & 1771.514 \\
	& COS/G160M & LB6B02050 &2010-08-22 18:52:45 & 1396.160& 1434.871 & 1795.561 \\
DN~Tau$^3$	& COS/G160M & LB6B04030 &2011-09-10 18:14:50& 965.120 & 1435.325  & 1796.015 \\
	& COS/G160M & LB6B04040 &2011-09-10 18:34:41 & 1387.136 & 1387.879 & 1748.477\\
	& COS/G160M & LB6B04050 &2011-09-10 19:50:39 & 1387.168 & 1411.111 & 1771.74\\
UZ~Tau$^4$ & COS/G160M & LBH208020 &2011-03-06 22:48:19 & 1944.288& 1384.094 & 1748.243 \\
 
\hline
\end{tabular}

\begin{tabular}{ll}
$^1$ & Observations obtained under HST proposal 8627 on "Testing the theories of wind/jet
production in YSOs" with PI, Nuria Calvet \\
$^2$ & Observations obtained under HST proposal 11533 on "Accretion flows and winds of
PMS stars" with PI, James Green \\
$^3$ & Observations obtained under HST proposal 11616 on "The disks, accretion and
outflows of TTSs" with PI, Gregory Herczeg \\
$^4$ & Observations obtained under HST proposal 12161 on
"Accretion in close pre-main sequence binaries" with PI, David Ardila \\
\end{tabular}
\end{center}
\end{tiny}
\end{table*}

\newpage
\begin{table*}

\begin{tiny}
\begin{center}
\caption{COS/G130M observations of the N~V line.\label{tab:table2}}
\begin{tabular}{lllllll}
\hline\hline
 Star &  Observation  &  Start Time          & Exposure   & Spec. Initial & Spec. Final \\
      &  ID           & (yyyy-mm-dd hh:mm:ss)& Time (sec) & Wavelength (\AA )& Wavelength (\AA )\\
\hline
SU~Aur$^1$	& LB6B11040 &2011-03-25 09:19:35 & 894.048 & 1132.513 & 1433.611 \\
	& LB6B11050 &2011-03-25 09:39:54 & 894.016 & 1171.264 & 1471.991 \\
LkCa~19$^1$ & LB6B28010 &2011-03-30 23:21:36 & 972.032 & 1132.687 & 1433.865 \\
	& LB6B28020 &2011-03-31 00:57:29 & 972.032 & 1171.018 & 1472.086 \\
GM~Aur$^1$ 	& LB6B01020 &2010-08-19 15:30:01 &1064.000 & 1173.883 & 1472.056 \\
	& LB6B01010 &2010-08-19 14:07:11 &1064.032 & 1135.175 & 1434.068 \\
RW~Aur~A$^1$& LB6B15040 &2011-03-25 04:20:18 & 881.920 & 1134.524 & 1433.479 \\
	& LB6B15050 &2011-03-25 04:40:25 & 882.016 & 1171.212 & 1471.940 \\
HN~Tau$^1$  & LB6B09040 &2010-02-10 17:01:25 & 2862.368& 1172.119 & 1472.816 \\
	& LB6B09050 &2010-02-10 18:37:17 & 2862.368& 1133.578 & 1434.835 \\
UX~Tau~A$^1$& LB6B13010 &2010-12-12 13:21:36 & 814.016 & 1132.756 & 1433.955 \\
	& LB6B13020 &2010-12-12 13:40:33 & 813.976 & 1178.443 & 1472.069 \\
HBC~427$^1$ & LB6B26010 &2011-03-29 23:42:24 &1116.000 & 1132.765 & 1433.914 \\
	& LB6B26020 &2011-03-30 01:11:12 &1007.040 & 1171.207 & 1472.055 \\
AA~Tau$^1$  & LB6B07040 &2011-01-06 23:15:54 &2844.320 & 1171.275 & 1471.974 \\
	& LB6B07050 &2011-01-07 00:51:43 &2844.352 & 1132.673 & 1433.841 \\
DR~Tau$^1$  & LB6B14040 &2010-02-15 12:31:03 &851.936  & 1135.968 & 1434.705 \\
	& LB6B14050 &2010-02-15 13:40:31 &852.032  & 1171.490 & 1467.855 \\
LkCa~4$^1$  & LB6B27010 &2011-03-30 06:05:02 &1152.000 & 1132.813 & 1434.013 \\
	& LB6B27020 &2011-03-30 06:29:21 &1152.000 & 1171.085 & 1472.155 \\
V836~Tau$^1$& LB6B06040 &2011-02-05 07:48:28 &2852.384 & 1171.243 & 1471.854 \\
	& LB6B06050 &2011-02-05 09:24:21 &2852.384 & 1132.642 & 1433.753 \\
DE~Tau$^1$ 	& LB6B08010 &2010-08-20 12:23:06 &1033.984 & 1132.918 & 1431.810 \\
	& LB6B08020 &2010-08-20 13:50:31 &1033.984 & 1172.592 & 1472.092 \\ 
DF~Tau$^2$  & LB3Q02030 &2010-01-11 07:10:07 &1000.192 & 1136.215 & 1429.787 \\
	& LB3Q02040 &2010-01-11 08:20:23 &1001.216 & 1145.843 & 1439.456 \\
	& LB3Q02050 &2010-01-11 08:40:31 &1410.208 & 1155.476 & 1449.112 \\
	& LB3Q02060 &2010-01-11 09:56:16 &1416.192 & 1165.040 & 1458.662 \\
IP~Tau$^1$ 	& LB6B05040 &2011-03-21 09:54:41 &1439.360 & 1139.996 & 1433.623 \\ 
	& LB6B05050 &2011-03-21 11:08:05 &1852.384 & 1170.797 & 1467.013 \\ 
DM~Tau$^1$ 	& LB6B02010 &2010-08-22 15:19:45 &1812.384 & 1171.278 & 1472.153 \\ 
	& LB6B02020 &2010-08-22 16:50:38 &1646.336 & 1132.819 & 1434.034 \\
DN~Tau$^1$  & LB6B04010 &2011-09-10 15:38:59 &1252.320 & 1171.707 & 1472.443 \\
        & LB6B04020 &2011-09-10 16:52:59 &1651.360 & 1133.030 & 1434.215 \\
\hline

\end{tabular}

\end{center}
\begin{tabular}{ll}
$^1$ & Observations obtained under HST proposal 11533 on "Accretion flows and winds of
PMS stars" with PI, James Green \\
$^2$ & Observations obtained under HST proposal 11616 on "The disks, accretion and
outflows of TTSs" with PI, Gregory Herczeg \\

\end{tabular}
\end{tiny}
\end{table*}

\clearpage
\begin{table*}
\begin{center}
\begin{scriptsize}
\caption{Main properties of the TTSs\label{tab3}}
\begin{tabular}{llllllll}
\tableline\tableline
Object & Spectral & L$_*$ \tablenotemark{(a)}& A$_V$ \tablenotemark{(a)}& Vsin(i)\tablenotemark{(f)} & Age\tablenotemark{(a)} & Period 	& L$_X$   \\
	&Type     &(L$_{\odot}$) 	     &(mag.) 			& (km/s) & $\log \tau$ (yr) & (d) 
&10$^{30}$ erg s$^{-1}$   \\
\tableline
SU  Aur    &G1  &7.8  				&0.90 	&66.2   &6.80$\pm$ 0.08	&3.5     & 11.641  \\
T  Tau     &K0  &7.8  				&1.8  	&20.1  	&\nodata       	&2.8     & 9.395    \\
LkCa 19    &K0  &1.56\tablenotemark{(b)}	&  0  	&19     &7.17$\pm$ 0.01	&2.24    & \nodata \\
GM  Aur    &K3  &1.2  				&1.1  	&12.4  	&6.9$\pm$ 0.2   &12      & 0.69\tablenotemark{(d)} \\
RW  Aur    &K4  &1.72\tablenotemark{(b)}	&0.32 	&17.2  	&7.20$\pm$ 0.11 &5.4     & \nodata \\
HN  Tau    &K5  &0.7\tablenotemark{(c)} 	&0.52 	&52.8	&\nodata 	&\nodata & 0.18\tablenotemark{(d)} \\
UX Tau	   &K5  &1.29\tablenotemark{(b)}	&0.0  	&26     &6.43$\pm$0.17  &\nodata & \nodata  \\
HBC 427    &K5Bin& 2			 	&0.6  	&\nodata&\nodata   	&\nodata &6.3      \\
AA  Tau    &K7  &1.1  				&1.4  	&11	&\nodata 	&8.2     &1.039    \\
DR  Tau    &K7  &1.7  				&1.0  	&$<10.0$  &5.92$\pm$ 0.2 	&7.3     \\
LkCa 4     &K7  &0.73\tablenotemark{(b)}	&0.69 	&26.1   &6.43$\pm$ 0.25 &3.38    & 1.99\tablenotemark{(d)}  \\
V836 Tau   &K7  &0.8 \tablenotemark{(d)}	&1.1    &$<15.0$  &\nodata      	&6.76    \\
DE  Tau    &M0  &1.2  				&1.1  	&10.0  	&\nodata    	&7.6     & 0.025\tablenotemark{(d)}  \\
IP  Tau    &M0  &0.7  				&0.9  	&11.0  	&6.6$\pm$ 0.2  	&3.3     &\nodata     \\
DF  Tau    &M1,M3.5&0.47,0.53 \tablenotemark{(e)}&0.04	&16.1  	&6.28$\pm$ 0.17 &8.5     & \nodata  \\
DM  Tau    &M1  &0.35 				&0.6  	&10	&6.87$\pm$ 0.34 &\nodata &1.181    \\
DN Tau     & M1 &1.2    			&0.8 	&8.1	&6.15$\pm$ 0.11 &6.6     &1.072    \\
UZ  Tau    &M1  &0.31 				&1.0  	&15.9  	&\nodata   	&\nodata &0.736    \\
\tableline
\end{tabular}
\tablenotetext{(a)}{ $L_*$, A$_V$ and age as in GdCMA2012 unless otherwise indicated.}.
\tablenotetext{(b)}{Bertout et al. 2007.}
\tablenotetext{(c)}{ $L_*$ in Table~1 from Ingleby et al. 2009 too small. Kenyon \& Hartmann 1995 used instead.}
\tablenotetext{(d)}{Hao Yang et al. 2012}
\tablenotetext{(e)}{  Luminosities of DF~Tau~A and DF~Tau~B are provided (from Bertout et al. 2007).}
\tablenotetext{(f)}{Clarke \& Bouvier 2000}

\end{scriptsize}
\end{center}
\end{table*}

\clearpage
\begin{table*}
\begin{center}
\begin{scriptsize}
\caption{Measurements and data \label{tab3}}
\begin{tabular}{llllll}
\tableline\tableline
Object & F(He~II) & F$_{1666}$(O~III])\tablenotemark{(a)} & F (N~V)\tablenotemark{(b)} & F$_{1338}$ (H$_2$) & F$_{1489.636}$ (H$_2$) \\
	& \multicolumn{5}{c}{(10$^{-14}$~erg~s$^{-1}$~cm$^{-2}$)} \\
\tableline
SU  Aur    & 2.05$\pm$ 0.13 & 0.46$\pm$ 0.18 $^w$  & 0.53$\pm$ 0.05 $^c$ & 2.82 $\pm$ 0.14 & 0.41 $\pm$ 0.04\\
T  Tau     & 4.87$\pm$ 0.96 & 1.14$\pm$ 0.41 $^w$  & \nodata             & \nodata         & \nodata        \\
LkCa 19    & 1.43$\pm$ 0.10 & \nodata              & 0.22$\pm$ 0.03 $^n$ & \nodata         & \nodata        \\
GM  Aur    & 6.22$\pm$ 0.19 & \nodata              & 2.17$\pm$ 0.22 $^n$ & 8.76 $\pm$ 0.67 & 1.08 $\pm$ 0.06\\
RW Aur     & 0.66$\pm$ 0.12 & 2.10$\pm$ 0.43 $^s$  & 0.66$\pm$ 0.11 $^b$ & 6.49 $\pm$ 0.91 & 3.26 $\pm$ 0.15\\
HN  Tau    & 1.33$\pm$ 0.26 & 1.24 $\pm$ 0.26 $^i$ & 0.27$\pm$ 0.03 $^c$ & 4.45 $\pm$ 0.23 & 1.04 $\pm$ 0.07\\
UX Tau	   & 2.25$\pm$ 0.12 & \nodata              & 0.88$\pm$ 0.08 $^c$ & 4.29 $\pm$ 0.21 & 0.85 $\pm$ 0.03\\
HBC 427    & 0.77$\pm$ 0.10 & \nodata              & 0.08$\pm$ 0.02 $^n$ & \nodata         & \nodata        \\
AA  Tau    & 2.93$\pm$ 0.06 & \nodata              & 0.14 $\pm$ 0.01$^c$ & 9.12 $\pm$ 0.24 & 1.16 $\pm$ 0.05\\
DR  Tau    & 0.93$\pm$ 0.15 & 2.14 $\pm$ 0.49 $^s$ & 0.25 $\pm$ 0.05$^c$ & 2.33 $\pm$ 0.34 & 0.52 $\pm$ 0.09\\
LkCa 4     & 0.54$\pm$ 0.06 & \nodata              & 0.07 $\pm$ 0.01$^n$ & \nodata         & \nodata \\
V836 Tau   & 0.44$\pm$ 0.03 & \nodata              & 0.07 $\pm$ 0.01$^c$ & 0.45 $\pm$ 0.05 & 0.15 $\pm$ 0.01\\
DE  Tau    & 0.87$\pm$ 0.16 & 0.27 $\pm$ 0.11 $^i$ & 0.35 $\pm$ 0.05$^b$ & 2.76 $\pm$ 0.28 & 0.45 $\pm$ 0.05\\
IP  Tau    & 1.33$\pm$ 0.04 & \nodata              & 0.08 $\pm$ 0.02$^c$ & 0.17 $\pm$ 0.05 & 0.13 $\pm$ 0.02\\
DF  Tau    & 7.32$\pm$ 0.20 & 1.42 $\pm$ 0.47 $^s$ & 1.07 $\pm$ 0.07$^b$ & 15.9$\pm$ 0.6   & 0.94 $\pm$ 0.07\\
DM  Tau    & 7.28$\pm$ 0.18 & \nodata              & 0.68 $\pm$ 0.10$^c$ & 6.72 $\pm$ 0.31 & 0.61 $\pm$ 0.03\\
DN Tau     & 6.53 $\pm$0.06 & \nodata              & 0.88 $\pm$ 0.08$^n$ & 0.88 $\pm$ 0.13 & 0.15 $\pm$ 0.04\\
UZ  Tau\tablenotemark{(c)}    & 1.06$\pm$ 0.34& 0.43 $\pm$ 0.33 & \nodata             & \nodata         & 0.55 $\pm$ 0.02\\
\tableline
\end{tabular}
\tablenotetext{(a)}{$w, i$ and $s$ stand for weak, intermediate and strong line, respectively. The flux in the 1660.802~\AA\ line has only been measured for strong sources. F$_{1660}$ is equal to $(0.37 \pm 0.32) \times 
10^{-14}$~erg~s$^{-1}$~cm$^{-2}$ for DF~Tau,  $(0.34 \pm 0.23) \times 
10^{-14}$~erg~s$^{-1}$~cm$^{-2}$ for DR~Tau and $(0.46 \pm 0.23) \times 
10^{-14}$~erg~s$^{-1}$~cm$^{-2}$ for RW~Aur. }.
\tablenotetext{(b)}{The line flux may be affected by the H$_2$ feature. A quality flag has been inserted. Fluxes flagged with $^b$ (strong blends), $^c$ (clean blends where the H$_2$ and N~V contributions can be separated) and $^n$ (no H$_2$ emission is detected). }
\tablenotetext{(c)}{UZ Tau is a close binary and observations during various phases are available. H$_2$ lines flux is constant but He~II and O~III] fluxes are variable. The values provided
in the table correspond to the mean and standard deviation of the measured fluxes during the cycle. During the cycle the He~II and O~III] fluxes vary by a factor of 3.}
\end{scriptsize}
\end{center}
\end{table*}

\clearpage
\begin{table}
\begin{center}
\begin{scriptsize}
\caption{UV lines profiles dispersion}
\begin{tabular}{lllll}
\tableline\tableline
Stars & $\sigma$(H$_2$ 1339\AA ) &$\sigma$(H$_2$ 1489\AA ) & $\sigma$(N~V 1238 \AA ) & $\sigma$(He~II) \\
& (km~s$^{-1}$) & (km~s$^{-1}$) & (km~s$^{-1}$) & (km~s$^{-1}$) \\
\tableline
       AATau   &    52.38   &    33.32  &     89.66  &     40.19 \\
       DETau   &    39.61   &    26.72  &     79.97  &     68.05 \\
       DFTau   &    52.38   &    36.71  &     116.3  &     48.66 \\
       DMTau   &    48.51   &     21.8  &     138.1  &     88.85 \\
       DNTau   &    48.51   &    39.81  &     36.35  &      39.7 \\
       DRTau   &     34.3   &    34.49  &     70.27  &     70.34 \\
       GMAur   &    65.69   &    30.85  &     123.6  &     62.25 \\
       HNTau   &    56.01   &    29.54  &     130.9  &     87.04 \\
       IPTau   &    19.81   &    33.32  &      72.7  &     32.58 \\
       SUAur   &    28.01   &    26.72  &       143  &      63.9 \\
       UXTau   &     34.3   &    23.55  &     99.35  &     58.41 \\
     V836Tau   &     34.3   &     32.1  &     67.85  &     33.26 \\
      LkCa19   &   \nodata  &   \nodata &     33.93  &     41.47 \\
       LkCa4   &   \nodata  &   \nodata &     38.77  &     36.05 \\
\tableline
\end{tabular}
\end{scriptsize}
\end{center}
\end{table}

\clearpage
\begin{table}
\begin{center}
\begin{scriptsize}
\caption{Two temperature fits to the XMM-Newton spectra from XEST \label{tab3}}
\begin{tabular}{llllll}
\tableline\tableline
Star & T$_1$ & T$_2$ & EM$_1$ & EM$_2$ & EM$_1$/EM$_2$ \\
     & (MK)  & (MK)  & 10$^{52}$ &   10$^{52}$ &  \\
\tableline
T~Tau & 4.52 & 23.65 & 33.86 & 57.85 & 0.585 \\
SU~Aur& 5.22 & 23.30& 87.01 & 32.45 & 2.681 \\
HBC~427& 9.04 & 28.06 & 14.98 & 15.07 & 0.994 \\
\tableline 
\end{tabular}
\end{scriptsize}
\end{center}
\end{table}

\clearpage
\begin{table}
\begin{center}
\begin{scriptsize}
\caption{Shift of the He~II narrow emission  component}
\begin{tabular}{lc}
\tableline\tableline
Stars & $\Delta \lambda = \lambda _{peak} - \lambda _0$ \\
      & (\AA ) \\
\tableline
       AATau   &  .175 \\
       DETau   &  .202 \\
       DFTau   &  .188 \\
       DMTau   &  .166 \\
       DNTau   &  .164 \\
       DRTau   &  .115 \\
       GMAur   &  .186 \\
       IPTau   &  .151 \\
       UXTau   &  .13 \\
     V836Tau   &  .166 \\
\tableline 
\end{tabular}
\end{scriptsize}
\end{center}
\end{table}


\begin{thebibliography}{}

\bibitem[Ardila et al.(2002)]{ardila2002} Ardila, D.~R., Basri, G., Walter, F.~M., et al. 2002, \apj, 566, 1100 
\bibitem[Ayres et al. 1995]{Ayres1995} Ayres, T. R., Fleming, T. A., Simon, T., et al., 1995, \apjs, 96, 223
\bibitem[Ayres 2005]{Ayres2005} Ayres, T.R., 2005, Proceedings of the 13th Cambridge Workshop on Cool Stars, Stellar Systems and the Sun, held 5-9 July, 2004 in Hamburg, Germany. Edited by F. Favata, G.A.J. Hussain, and B. Battrick. ESA SP-560, European Space Agency, p.419
\bibitem[Beristain et al. 2001]{Beristainetal2001}Beristain, G., Edwards, S., Kwan, J., 2001, \apj, 551, 1037
\bibitem[Bertout et al. 2007]{Bertout2007} Bertout, C., Siess, L. and Cabrit, S., 2007, Astronomy \& Astrophysics, 473, L21--L24
\bibitem[Brown et al. 1984]{brownetal84} Brown, A. de M. Ferraz, M. C. and Jordan, C., 1984, Royal Astronomical Society, 207, 831
\bibitem[Clarke and Bouvier 2000]{CK} Clarke, C.J., Bouvier, J., 2000, M.N.R.A.S., 319, 457
\bibitem[D'Antona and Mazzitelli(1997)]{dantona1997} D'Antona, F., \& Mazzitelli, I.\ 1997, \memsai, 68, 807 
\bibitem[Damiani et al. 1995]{damianietal1995} Damiani, F., Micela, G., Sciortino, S. et al. 1995, \apj, 446 331
\bibitem[Ferland 1996]{ferland1996} Ferland, G.,  J., 1996, {\it Hazy, a Brief Introduction to Cloudy}, University of Kentucky, Department of Physics and Astronomy Internal Report.
\bibitem[Fisher et al. 2008]{fischeretal2008} Fischer, W., Kwan, J., Edwards, S. et al., 2008, \apj, 687, 1117
\bibitem[France et al. 2012]{franceetal2012} France, K., Schindhelm, E., Herczeg, G. J. et al. 2012, \apj, 756, 171
\bibitem[Glassgold et al. 2000]{glassgoldetal2000} Glassgold,  A. E., Feigelson, E. D., Montmerle, T., 2000, Protostars and Planets IV (Book - Tucson: University of Arizona Press; eds Mannings, V., Boss, A.P., Russell, S. S.), p. 429
\bibitem[Gomez de Castro and Lamzin 1999]{gdcl1999} G\'omez de Castro, A.I. \& Lamzin, S., 1999, Monthly Notices of the R.A.S., 304, L41
\bibitem[Gomez de Castro and Verdugo(2001)]{aig2001} G{\'o}mez de Castro, A.~I., \& Verdugo, E.\ 2001, \apj, 548, 976 
\bibitem[Gomez de Castro and Verdugo(2003)]{aig2003} G{\'o}mez de Castro, A.~I., \& Verdugo, E.\ 2003, \apj, 597, 443 
\bibitem[Gomez de Castro et al. 2003]{gdcetal2003} G\'omez de Castro, A.I., Verdugo, E., Ferro-Font\'an, C., 2003,  The Future of Cool-Star Astrophysics: 12th Cambridge Workshop on Cool Stars, Stellar Systems, and the Sun (2001 July 30 - August 3), eds. A. Brown, G.M. Harper, and T.R. Ayres, (University of Colorado), p. 40-49.
\bibitem[G\'omez de Castro and Ferro-Font\'an 2005]{gdcff05} G\'omez de Castro, A.I., Ferro-Font\'an, C., 2005, \mnras , 362, 569
\bibitem[G{\'o}mez de Castro and  Verdugo(2007)]{aig2007} G{\'o}mez de Castro, A.~I., \& Verdugo, E.\ 2007, \apjl, 654, L91 
\bibitem[G{\'o}mez de Castro (2009)]{aig2009} G{\'o}mez de Castro, A.~I.\ 2009, \apjl, 698, L108 
\bibitem[G\'omez de Castro and von Rekowsky 2011]{gdcvR11} G\'omez de Castro, A.I., von Rekowsky, B., 2011, \mnras , 411, 849
\bibitem[Gomez de Castro and Marcos-Arenal 2012]{gdcma2012} G\'omez de Castro, A.I., Marcos-Arenal, P., 2012, \apj,  749, 190
\bibitem[Gomez de Castro 2013]{gdc2013} G\'omez de Castro, A.I. 2013, Planets, Stars and Stellar Systems, Editor-in-chief: Oswalt, T. D. McLean, I.S.; Bond, H.E.; French, L.; Kalas, P.; Barstow, M.A.; Gilmore, G.F.; Keel, W.C. (Eds.), Springer, ISBN 978-90-481-8817-8
\bibitem[Green et al. 2012]{greenetal2012} Green, J. C., Froning, C. S., Osterman, S. et al. 2012, \apj, 744, 60
\bibitem[G{\"u}del et al.(2007)]{gudel2007} G{\"u}del, M., Briggs, K.~R., Arzner, K., et al.\ 2007, \aap, 468, 353 
\bibitem[Guenther et al 1999]{guentheretal99} Guenther E.W., Lehman H., Emerson J.P. et al., \aap , 341, 768, 1999

\bibitem[Hartmann 2009]{hartmann2009} Hartmann, L., 2009, Protostellar Jets in Context, by Kanaris Tsinganos, Tom Ray, Matthias Stute. Astrophysics and Space Science Proceedings Series. Berlin: Springer, pp.23-32
\bibitem[Hartmann et al. 1982]{hartmannetal1982} Hartmann, L., Avrett, E. Edwards, S., 1982, \apj, 261, 279 
\bibitem[Herczed et al. 2002]{herczegetal2002} Herczeg, G. J., Linsky, J. L., Valenti, J. A., 2002, \apj, 572, 310
\bibitem[Hirth et al. 1997]{hetal1997} Hirth,G.A., Mundt, R., Solf, J., 1997, Astronomy \& Astrophysics Sup., 126, 437 
\bibitem[Hu\'elamo et al. 1998]{huelamoetal98} Hu\'elamo, N., G\'omez de Castro, A.I., Franqueira, M., 1998, Ultraviolet Astrophysics Beyond the IUE Final Archive,  ESA Publications Division, ESA SP,  413, 121
\bibitem[Ingleby et al. 2009]{Ingleby2009} Ingleby, L. Calvet, N., Bergin, E. et al., 2009, \apj, 703, L137
\bibitem[Johns-Krull et al. 1999]{jk1999} Johns-Krull, C. M., Valenti, J. A., Koresko, C., 1999, \apj, 516, 900
\bibitem[Johns-Krull et al.(2000)]{johnskrull2000} Johns-Krull, C.~M.,  Valenti, J.~A., \& Linsky, J.~L.\ 2000, \apj, 539, 815 
\bibitem[Johns-Krull et al.(2004)]{johnskrull2004} Johns-Krull, C.~M., Valenti, J.~A., \& Saar, S.~H.\ 2004, \apj, 617, 1204 
\bibitem[Johns-Krull 2007]{JK2007} Johns-Krull, C. M., 2007, \apj , 664 975
\bibitem[Johns-Krull 2009]{johnskrull2009} Johns-Krull, C.M., {\em Protostellar Jets in Context, Astrophysics and Space Science Proceedings Series}, 33, 2009
\bibitem[Kenyon and Hartmann 1995]{Kenyon1995} Kenyon, S. J. and Hartmann, L., 1995, \apss , 101, 117
\bibitem[Kivelson and Russell 1995]{kr1995} Kivelson, M.G., Russel, C.T., 1995, {\it Introduction to space physics}, Cambridge Univ. Press, p. 330
\bibitem[Kriss 2011]{k2011} Kriss, G.A. 2011, Improved Medium Resolution Line Spread Functions for
COS FUV Spectra, Technical Report
\bibitem[Lamzin 1999]{l1999} Lamzin, S., 1998, Astronomy Reports, 42, 322
\bibitem[Lemmens et al. 1992]{lemmensetal92} Lemmens, A. F. P., Rutten, R. G. M., Zwaan, C. 1992, \aap , 257, 671
\bibitem[Lima et al. 2010]{l2010} Lima, G.H.R.A., Alencar, S.H.P., Calvet, N. et al., 2010, \aap , 522, 104L
\bibitem[Linsky et al. 1982]{Linsky1982} Linsky, J. L., Bornmann, P. L., Carpenter, K. G. et al. 1982, \apj, 260, 670
\bibitem[Muzerolle et al. 1998]{metal1998} Muzerolle, J., Calvet, N., Hartmann, L., 1998, \apj, 492, 743
\bibitem[Muzerolle et al. 2001]{meta2001} Muzerolle, J., Calvet, N., Hartmann, L., 2001, \apj, 550, 994
\bibitem[Neuhauser et al. 1995]{Netal1995} Neuhauser, R., Sterzik, M.F., Schmidt, J.H.M.M. et al 1995, \aap , 297, 391
\bibitem[Orlando et al. 2009]{oet1l2009} Orlando, S., Sacco, G. G., Argiroffi, C. et al., 2009, \aap , 510, 71
\bibitem[Pedersen and Gomez de Castro 2011]{PGdC2011} Pedersen, A., G\'omez de Castro, A.I., 2011, \apj , 740, 77
\bibitem[Penston and Lago 1983]{pl1983} Penston, M. \& Lago, M.T.V.T., 1983, Monthly Notices of the R.A.S., 202, 77
\bibitem[Petrov et al. 2011]{petal2011} Petrov, P.P., Gahm, G., F., Stempels, H.C. et al., 2011, \aap , 535, 6
\bibitem[Romanova et al. 2004]{retal2004} Romanova, M. M., Ustyugova, G. V., Koldoba, A. V. et al., 2004, \apj, 610, 920
\bibitem[Romanova et al. 2012]{retal2012} Romanova M. M., Ustyugova, G. V., Koldoba, A. V. et al. 2012, Monthly Notices of the R.A.S., 421, 63
\bibitem[Sachs 1982]{s1982} Sachs, L., 1982, Applied Statistics, a handbook of techniques, Springer Series in Statistics, Springer-Verlag, NY
(ISBN:0-387-90976-1)
\bibitem[Siess et al.(2000)]{siess2000} Siess, L., Dufour, E., \& Forestini, M.\ 2000, \aap, 358, 593 
\bibitem[Valencic et al.(2004)]{valencic2004} Valencic, L.~A., Clayton, G.~C., \& Gordon, K.~D.\ 2004, \apj, 616, 912 
\bibitem[Yang et al 2012]{yatal2012} Yang, H., Herczeg, G. J., Linsky, J. L., et al, 2012, \apj, 744, 121


\end{thebibliography}
\end{document}